\documentclass[tbtags,floatfix,aps,prb,twocolumn,showpacs,superscriptaddress,groupedaddress]{revtex4-2}
\usepackage{graphicx, color}  
\usepackage{dcolumn}   
\usepackage{bm} 
\usepackage{amssymb}   
\usepackage{float}
\usepackage{hyperref}
\usepackage{xcolor}
\definecolor{link}{rgb}{0.1,0.1,0.9}
\definecolor{link1}{rgb}{0.1,0.1,0.9}
\hypersetup{colorlinks=true,citecolor=link,linkcolor=link, urlcolor=link1}

\usepackage{epstopdf}
\usepackage[tbtags]{amsmath}
\usepackage{amssymb}
\usepackage{amsmath}
\usepackage{longtable}
\usepackage{mathtools}

\begin{document}
	\preprint{APS/123-QED}



\title{Graphene Straintronics by Molecular Trapping}
\author{Pawan Kumar Srivastava$^1$, Vedanki Khandelwal$^{2,3}$, Ramesh Reddy$^4$, Kartick Tarafder$^4$ and Subhasis Ghosh$^2$}

\email{subhasis.ghosh.jnu@gmail.com}
\affiliation{$^1$School of Mechanical Engineering, Sungkyunkwan University, Suwon 16419, South Korea}
\affiliation{$^2$School of Physical Sciences, Jawaharlal Nehru University, New Delhi 110067, India}
\affiliation{$^3$Govt. Adarsh Girls College, Sheopur, Madhya Pradesh 476337, India}
\affiliation{$^4$Department of Physics, National Institute of Technology Karnataka, Mangalore 575025, India}

\date{\today}
\begin{abstract}
Here, we report on controlling strain in graphene by trapping molecules at the graphene-substrate interface, leveraging molecular dipole moments. Spectroscopic and transport measurements show that strain correlates with the dipole moments of trapped molecules, with a dipole range of 1.5 D to 4.9 D resulting in a 50-fold increase in strain and a substantial rise in the residual carrier density. This has been possible by charge transfer between graphene and trapped molecules, altering the C=C bond length, and causing biaxial strain. First-principles density functional theory calculations confirm a consistent dependence of bending height on molecular dipole moments.
\end{abstract}

\maketitle

Graphene straintronics provides a way to modify both the electronic and mechanical properties of graphene, making it highly versatile for use in new and emerging technologies \cite{ghosh2023graphene,
hou2024strain,solomenko2023straintronics}. Recent discoveries highlight strain as a crucial parameter for manipulating the electronic and optical characteristics of two-dimensional layered materials \cite{peng2020strain,liu2014strain,mennel2018optical,yang2021strain,lee2012optical}. Controlling strain in 2D materials is significantly more feasible than in their 3D counterparts, due to their atomic-scale thickness. Graphene, a 2D layered material, exhibits a remarkable strain limit due to its extraordinary tensile strength and provides an effective means to fine-tune its electronic properties \cite{lee2008measurement,choi2010effects}. Strain-induced modulation in graphene leads to the emergence of novel phenomena, such as opening of an energy band-gap \cite{postorino2020strain}, non-trivial quantum phase transition in graphene superlattices \cite{parker2021strain} and emergence of large pseudo-magnetic fields, reaching up to 300 Tesla \cite{settnes2016pseudomagnetic,levy2010strain,zhang2022strain,banerjee2020strain}. Generally, ripples and corrugations, the dominant source of strain fluctuations (SFs), are present in substrate-supported graphene \cite{locatelli2010corrugation,zihlmann2020out,couto2014random,wang2020mobility}. The reproducibility and quality of graphene-based devices remain significantly compromised due to the lack of experimental control over SFs. To address this issue, several methods have been explored to control strain in graphene. These include transferring graphene onto flexible or piezoelectric substrates \cite{yang2021strain,wang2020mobility,ni2008uniaxial,roldan2015strain,mohiuddin2009uniaxial}, performing nanoindentation on suspended graphene \cite{lee2008measurement,won2022raman}, and selecting substrates with a thermal expansion coefficient different from that of 2D materials to induce strain \cite{yang2021strain,ahn2017strain}. There are problems with these methods. Firstly, in most cases strain is uncontrolled. Secondly, the physics behind strain engineering is either not known or poorly understood. Furthermore, there has been limited success in developing an effective method to precisely control the strain in substrate-supported graphene.

Straintronics of graphene is an important problem for both fundamental research and future technologies which are addressed in this letter. Firstly, we report on controlling strain in substrate-supported graphene by trapping molecules at the graphene-substrate interface. Secondly, we reveal the underlying physics governing the microscopic origin of the strain in graphene. The trapped molecules induce biaxial strain in graphene by altering the C=C bond length through charge transfer between graphene and the molecules. Notably, we show that strain can be increased by almost 50 times using molecules with high dipole moment. Shifts in the Dirac point ($V_{Dirac}$), minimum conductivity ($\sigma_o$), and residual carrier density ($n_o$) with the dipole moment of the trapped molecules confirm the charge transfer and resultant increase in the C=C bond length. These experimental observations have been corroborated by first-principle density functional theory (DFT) calculations.

Molecular trapping was facilitated through liquid phase exfoliation of graphene layers described elsewhere \cite{srivastava2017controlled}. Graphene dispersion after sonication and centrifugation, was wet transferred to the $SiO_2$ substrates and subsequently annealed in a vacuum above the boiling point of respective solvents. More details about sample preperation are given in supplemental material (SM), section S1 \cite{SM} (see also Refs. \cite{berry2013impermeability, sun2020limits, kresse1996efficiency, perdew1992atoms, perdew1996generalized,
goumans2007structure, srivastava2017relativistic,
bruna2014doping, ferrari2013raman, barrios2013pseudo} therein). Since graphene acts as an impermeable membrane \cite{joshi2014precise,tsetseris2014graphene,bunch2008impermeable}, solvents at the interface will be trapped and will not be able to evaporate even at elevated temperatures (for details, see SM, section S2 \cite{SM}). Molecules trapped at graphene-substrate interface was confirmed using confocal Raman and infrared spectroscopy which were performed using WITec GmbH Alpha confocal Raman microscope and Varian-7000 UMA-600 IR microscope, respectively. Details of device fabrication (section S4), confirmation of monolayer graphene (section S5) and DFT calculations (section S6) are given in SM \cite{SM}.
\begin{figure}
\centering
	\includegraphics[width=8.5 cm]{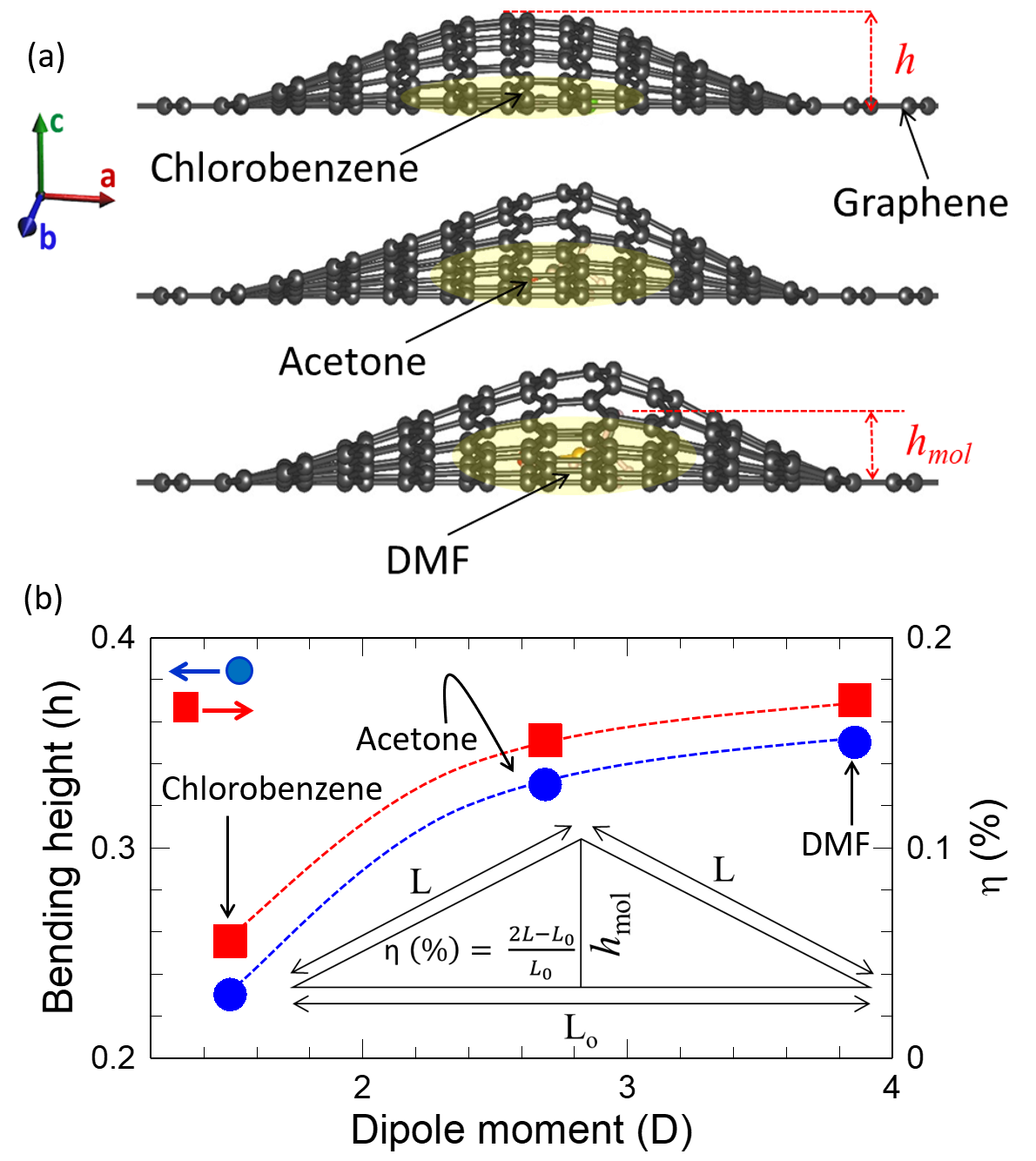}
\caption{\textbf{Strain due to molecular trapping at the graphene-substrate interface} \textbf{(a)} First principle density functional theory calculation showing bending in graphene basal plane because of molecular trapping at the graphene-substrate interface. Carbon atoms in graphene and trapped molecules are labeled. The bending height of graphene and molecules dimension in the c-axis is denoted by $h$ and $h_{mol}$, respectively. \textbf{(b)} Bending in graphene as a function of the dipole moment of trapped molecules, estimated from DFT calculations (graphene dimension for this calculation is 3 nm x 3 nm; see SM \cite{SM}). Bending height which results in mechanical strain in graphene, exceeds the c-axis molecular dimension and substantially depends on the polarity of trapped molecules. Strain $\eta$, calculated using the bending height and graphene dimension, as a function of Dipole moment is also given. The model to calculate strain is also schematically illustrated in the inset. Schematic illustration of strain in graphene due to molecular trapping is given in SM \cite{SM} (see Fig. S5)}.
	\label{DFT calulation}
\end{figure}

Figure~\ref{DFT calulation} shows the predicted strain evolution in graphene through molecular trapping at the graphene-substrate interface, based on our DFT calculations (details in the SM, section S6 \cite{SM}). Fig.~\ref{DFT calulation}a shows the relaxed structure obtained through DFT calculations of graphene on top of the trapped molecules, as labeled. Three major observations could be drawn from the relaxed structures: (i) the bending height $h$ in graphene increases monotonically with the dipole moment of the trapped molecules, as illustrated in Fig.~\ref{DFT calulation}b, (ii) $h$ exceeds the trapped molecule height in the c-direction, $h_{mol}$ (discussed later), and (iii) C=C bonds in graphene appear to be stretched in the vicinity of the trapped molecules with large dipole moments. Notably, as illustrated in Fig.~\ref{DFT calulation}b, the bending of graphene due to trapped molecules suggests that strain is generated through molecular trapping at the interface. The increase in $h$ values, or strain ($\eta$), with the dipole moment of trapped molecules, can be explained by the charge transfer between these molecules and graphene (details of charge transfer mechanism are given in SM, section S5 \cite{SM}). This makes the dipole moment a crucial factor in effectively tuning the $\eta$. Polar molecules, like dimethylformamide (DMF) or propylene carbonate (PC), tend to either withdraw or donate electrons to graphene. We have already shown that highly polar molecules transfer a greater amount of charge to graphene \cite{khandelwal2023revealing,srivastava2017controlled}. This process affects the electronic environment of the graphene, which in turn influences the interatomic forces and the equilibrium bond lengths, leading to strain \cite{dong2009doping}. This observation is akin to the electric field-driven strain observed in ferroelectric crystals, where electrostatic doping-induced asymmetric displacement in crystal dimensions has been reported \cite{ren2004large}. Furthermore, we want to highlight that the strain in graphene could arise from factors such as lattice mismatch, wrinkles, dislocations, and doping due to ambient oxygen or during the growth procedure. However, we have ruled out these effects using experimental results and valid arguments (details are provided in SM, section S3 \cite{SM}). 

According to first-principles calculation (see Fig. S4 \cite{SM}), adjusting the dipole moment of trapped molecules, ranging from 1.5D (chlorobenzene) to 4.9D (PC), allows manipulation of biaxial strain from 0.06\% to 0.3\% (see Fig.~\ref{DFT calulation}). We have chosen these organic molecules to control strain, facilitating both the exfoliation of graphene and the trapping of molecules at the graphene-substrate interface in various molecular environments. We have shown that graphene can be doped by charge transfer from (to) the graphene to (from) the molecule and the degree of the charge transfer only depends on the dipole moment of the molecule \cite{srivastava2017controlled}. As a proof of concept, we proceeded with a customized experimental setup where we used liquid phase exfoliation of highly oriented pyrolytic graphite (HOPG) to isolate graphene, which provides a well-developed toolbox to realize molecular trapping at the graphene-substrate interface (see SM, section S1 and S3 \cite{SM}). Such exfoliation is facilitated by the transfer of graphene layers from the solution to $SiO_2$/Si ($n^{++}$), where the residual molecules from the exfoliation process remain (trapped) at the graphene-substrate interface. For this purpose, we used various molecules with varying dipole moments, such as chlorobenzene (1.5D), acetone (2.69D), DMF (3.68D), and PC (4.9D) to manipulate strain in the graphene placed on $SiO_2/Si$ substrates.
\begin{figure*}
	\includegraphics[width= 16cm] {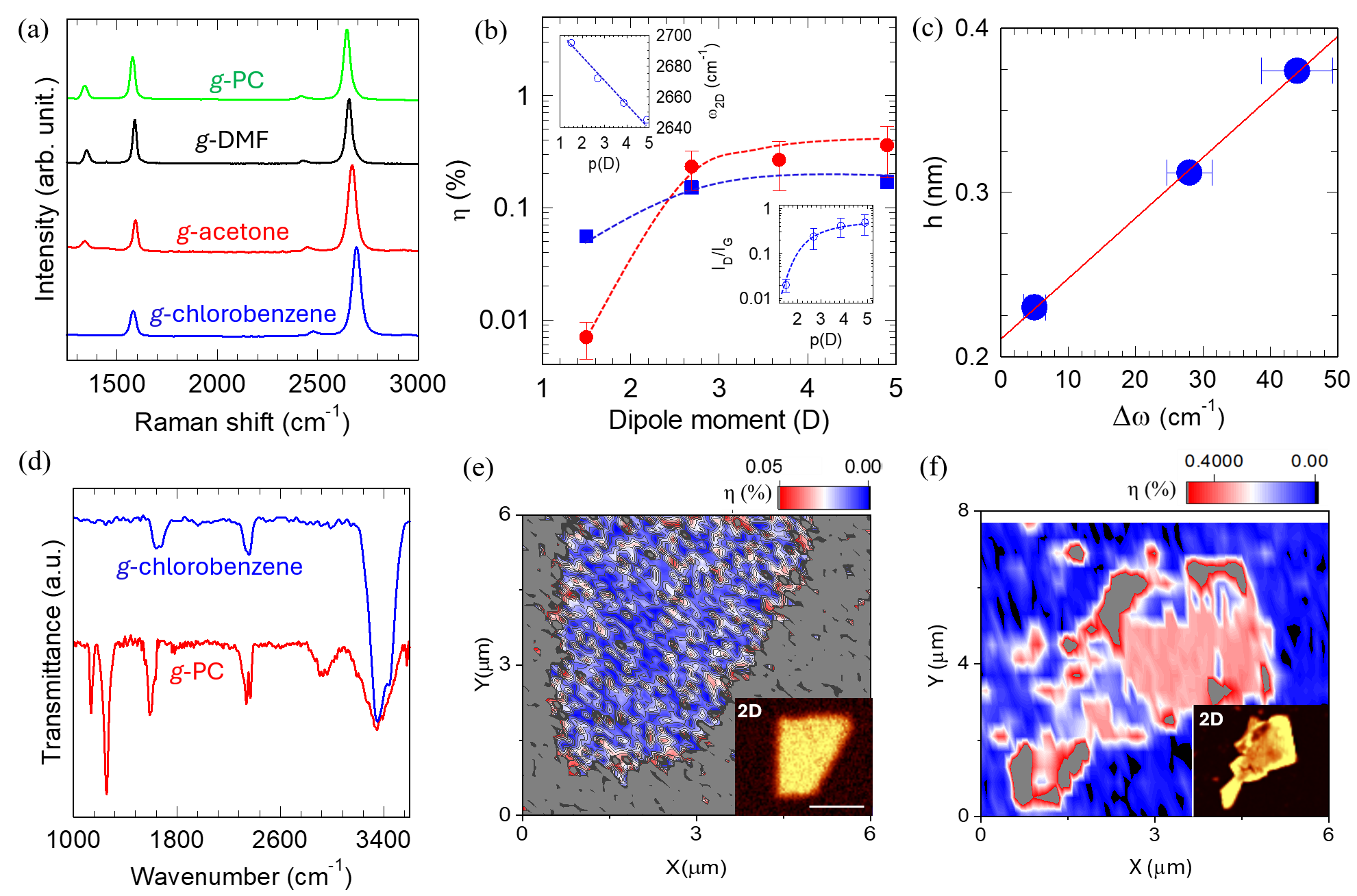}
	\caption{\textbf{Probing strain with Raman spectroscopy (a)} Raman spectra of graphene layers for various trapped molecules as labeled. \textbf{(b)} Strain (calculated: filled rectangle, and experimental: filled circles with error bar; given in the main panel), 2D Raman peak position (upper inset), and normalized Raman D peak ratio (lower inset) in graphene as a function of dipole moment of trapped molecules. Monotonic redshift in the $\omega_{2D}$ with dipole moment indicates strained graphene. Moreover, similar evolution of strain $\eta$, and D peak with dipole moment suggests a strong correlation between these two parameters. Symbols are the data points and dashed lines are the guide to eyes. \textbf{(c)} Bending height as a function of Raman shift in 2D peak position. The arrow represents the intercept on Y-axis ($h \neq 0$ as $\Delta\omega \to 0$) suggesting residual strain due to extrinsic effects. Symbols are the data points and solid line is the linear fit. \textbf{(d)} FTIR of g-chlorobenzene and g-PC. 2D map of spatial distribution of strain in case of \textbf{(e)} g-chlorobenzene and \textbf{(f)} g-DMF. Respective 2D Raman maps for 2D peaks are also given. Scale bar for Raman maps: 3 $\mu m$}
	\label{Raman spectra}
\end{figure*}
Fig.~\ref{Raman spectra}a shows Raman spectra of graphene layers with various trapped molecules at the interface. An increase in the dipole moment of the trapped molecules causes (i) a red shift in the Raman 2D peak and (ii) enhanced Raman D peak intensity. These two observations are closely linked to the molecular trapping and resultant strain in graphene. The red (or blue) shift of the 2D peak in the Raman spectra of graphene, provides insights into either electron or hole doping in graphene \cite{voggu2008effects,das2008monitoring}. It has been documented that when graphene is under uniaxial or biaxial tensile strain, Raman 2D peaks shift to lower frequency \cite{dong2009doping}. This red shift can be understood due to the elongation of the C=C bonds, this elongation weakens the bond strength and therefore, lowers their vibrational frequency. The monotonic red shift of the Raman 2D peak (see Fig.~\ref{Raman spectra}b, top inset) with the dipole moment of the trapped molecule indicates the increased electron density resulting from charge transfer from trapped molecules to graphene \cite{childres2013raman,das2008monitoring}. In undoped graphene, the $\pi$-bonding orbitals are fully occupied with electrons, while the anti-bonding $\pi^*$ orbitals are empty. The extra electrons, induced in graphene by charge transfer from the trapped molecules (n-type doping), populate the higher energy anti-bonding $\pi^*$ orbitals. This weakens the overall interactions between carbon atoms because anti-bonding orbitals have wavefunctions that oppose the bonding wavefunctions. As a result, the carbon atoms in the graphene lattice move slightly apart to reach a new equilibrium where the forces between them are balanced. Hence, the increase in electron density causes the expansion of graphene lattice, resulting in longer $sp^2$ C=C bond lengths \cite{kertesz1982change,pietronero1981bond}. The D-peak arises from the breathing modes of $sp^2$ rings and requires defects for activation. Elongated $sp^2$ C=C bonds in graphene naturally introduce disorder by disturbing the lattice periodicity, which activates the disorder-induced Raman D peak. Evidence that the D peak originates from elongated C=C bond comes from the fact that the ratio of intensities of the Raman D and G peaks $(I_D/I_G)$ increases monotonically with the dipole moment of trapped molecules (Fig.~\ref{Raman spectra}b, lower inset). If C=C bond breaking, substitutional doping during the trapping process, or uncontrolled disorder were responsible for the emergence of the Raman D peak, its monotonic dependence on the dipole moment would not occur. Hence, the monotonic increase in the Raman D peak results from the alterations of the $sp^2$ C=C bonds, reflecting increased strain (Fig.~\ref{Raman spectra}b, main panel), confirming a strong correlation between them. Here, we want to mention that the Raman modes are influenced by both doping and strain. However, in our case, these two factors are closely interrelated. Specifically, trapped molecules cause bending in the graphene and simultaneously induce charge transfer to graphene, which leads to C=C bond elongation. To substantiate this, we have carried out extensive analysis, with a detailed discussion provided in the SM, section S8 \cite{SM}.
We have deduced the strain from the red shift of Raman 2D peak using the expression \cite{metzger2010biaxial,androulidakis2015graphene,zabel2012raman} $\eta=\Delta{\omega}/2\omega_0\gamma_{2D}$ where $\Delta\omega$, $\omega_0$ and $\gamma_{2D}$ are shift in the Raman 2D peak, the Raman 2D peak position for clean graphene ($\sim2696$ $cm^{-1}$) and $Gr\Ddot{u}neisen$ parameter ($\sim2.6$) for biaxial strain \cite{won2022raman,mohiuddin2009uniaxial}, respectively. The approach used to calculate $\Delta$$\omega$ and the fitting of the Raman data which is used to quantify the central peak shift, are given in the SM, section S9 \cite{SM}. Fig.~\ref{Raman spectra}b summarizes strain evolution with dipole moment (p) of the trapped molecules. We have observed an enhancement in biaxial strain from 0.006\% (for p = 1.5D) to 0.3\% (p = 4.9D). Strain values extracted from Raman measurements were further reconciled with the model that considers the bending of graphene as a source of strain, as presented in the inset of Fig.~\ref{DFT calulation}b. In this model, we use bending height (estimated from DFT calculations) to extract mechanical strain expressed as $\eta (\%)=(2L-L_0)/L_0$ where $L_0$ and $L$ represent initial and final length scales, respectively. A similar trend for an increase in strain values in both cases corroborates molecular trapping at the interface as a primary source of biaxial strain in graphene. However, for lowest dipole moment molecule, $\eta= 0.006\%$ in experiments is ten-fold smaller compared to the theoretical prediction ($\eta = 0.06\%$). This can be explained using Fig.~\ref{Raman spectra}c, which shows the relationship between the theoretically derived bending height, $h$ and the Raman frequency shift ($\Delta{\omega}$) for graphene samples with trapped molecules of different dipole moments. As expected, we observe that the bending height increases with an increase in $\Delta{\omega}$. We find that there is a finite intercept on the Y-axis (h = 0.21 nm) even at $\Delta{\omega}$ = 0, which corresponds to a strain value of approximately 0.02\%. This indicates that in the low dipole moment regime, where charge transfer is minimal, the lower limit of strain will be due to the bending of graphene resulting from unavoidable corrugations or wrinkles present in all substrate supported graphene.\\
To further validate the monotonous increase in strain values with the dipole moment of trapped molecules, we have carried out Fourier transform infrared spectroscopy (FTIR) (Fig. \ref{Raman spectra}d). In graphene with chlorobenzene trapped at the interface (g-chlorobenzene), C=C (linked to graphene) and C-H vibrations (from molecule) can be observed at $1630$ $cm^{-1}$ and $3300$ $cm^{-1}$, respectively. However, for PC trapped at the interface (g-PC), in-plane C=C vibration is slightly red-shifted ($1610$ $cm^{-1}$) compared to g-chlorobenzene, indicating different bond lengths due to doping. In addition to C=C vibration in g-PC, additional peaks are also present at $1136$ $cm^{-1}$ and $1254$ $cm^{-1}$ (C-O stretching), and $2930$ $cm^{-1}$ and $3346$ $cm^{-1}$ (C-H stretching) which correspond to the trapped PC molecules and significantly contribute to the charge transfer from molecules to graphene. We have calculated C=C bond length using $Badger's$ rule \cite{badger1935relation,kaupp2017chemistry}, which relates bond length ($r_e$ in \AA) and force constant (k in Mdyne/cm) as follows-
\begin{equation}
     r_e = \Big({\frac{c_{ij}}{k}}\Big)^{1/3}+d_{ij}
\end{equation}
where k is directly calculated using wave number $(\nu)$ by the relation $k=4\pi^{2}c^2\nu^2\mu$ (c is the speed of light and $\mu$ is the reduce mass) and $c_{ij}$, $d_{ij}$ are fitting constants, in our case we have used $\sqrt[3]{c_{ij}} = 0.725$ and $d_{ij} = 0.68$ \cite{badger1935relation}. The calculated bond lengths are 1.4196$\r{A}$ and 1.4257$\r{A}$ for g-chlorobenzene and g-PC, respectively. The increase in bond length further supports the elongation of C=C bond length in molecules with high dipole moments. Due to negligible strain in g-chlorobenzene, we considered its C=C bond length as a reference and estimated a strain of $0.4\pm 0.05\%$ for g-PC that falls close to the value obtained from Raman and first-principles calculations. We have also examined the spatial distribution of strain across the graphene flakes. For g-chlorobenzene, the strain distribution appears to be minimal across the basal plane, as shown in Fig. ~\ref{Raman spectra}e. In contrast, g-DMF shows a significant strain distribution across the basal plane, as illustrated in Fig. \ref{Raman spectra}f. 

\begin{figure}
	\includegraphics[width=8.0 cm]{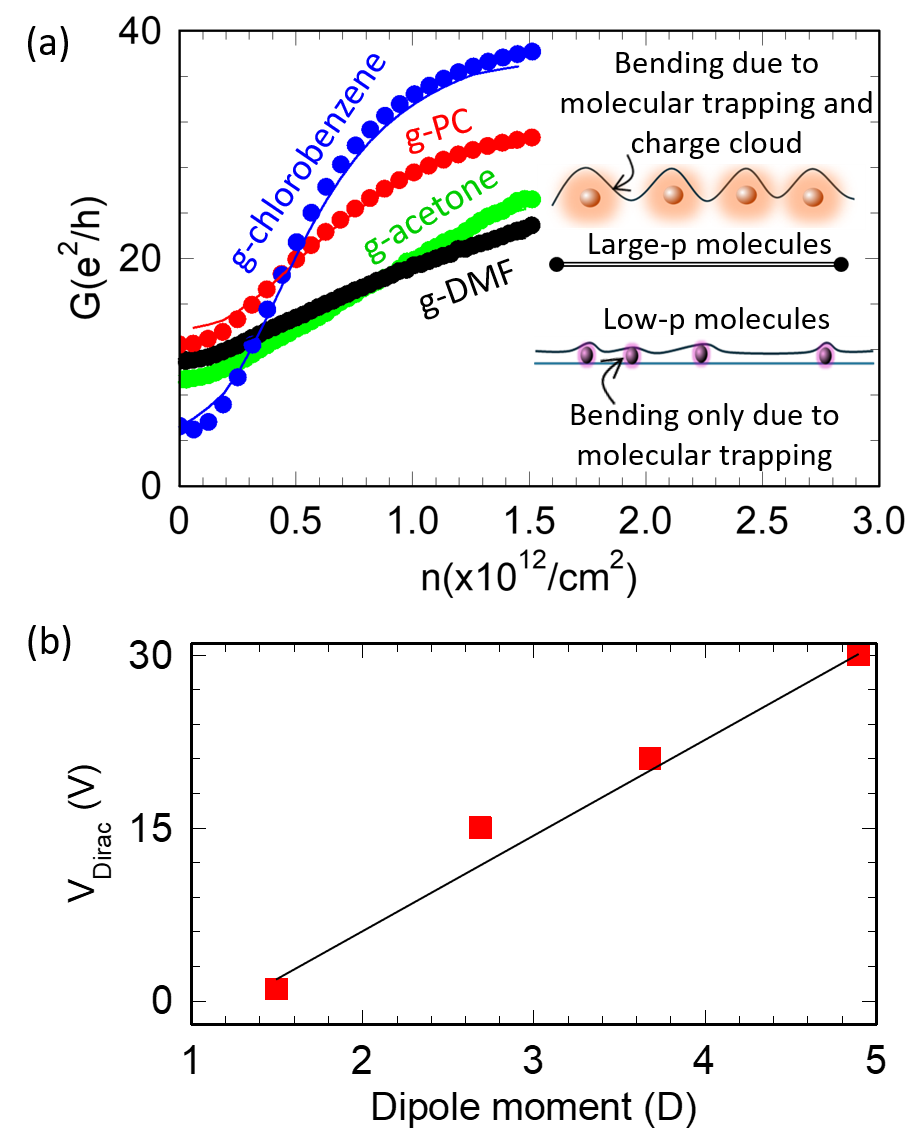}
	\caption{\textbf{Transport measurements (a)} Gate (carrier density, n) dependent differential conductance (electron side) of two terminal graphene FETs for various trapped molecules as labeled. Symbols are the data points and solid lines are fit according to Eq.\ref{eq2} (see text). The inset shows schematics for molecular trapping at the graphene-substrate interface. Molecules with low dipole moments have weaker interactions with graphene. In contrast, polar molecules interact more strongly with graphene and charge clouds (illustrated by shaded colored regions) near the graphene basal plane causing increased bending height \cite{khandelwal2023revealing} (see text for the explanation). \textbf{(b)} Variation in the Dirac point $V_{Dirac}$ as a function of the dipole moment of the trapped molecules at the interface.}
	\label{Transport measurement}
\end{figure}
Fig. \ref{Transport measurement}a shows two terminal conductance (G) as a function of gate-defined electron density (n) for various devices. Here n was calculated using expression \cite{das2008monitoring} $n=C_g(V_g-V_{Dirac})/e$ where $C_g$, $V_g$, $V_{Dirac}$ and e are gate capacitance, gate voltage, Dirac point and electronic charge, respectively. Minimum conductivity ($\sigma_0$) increases with the dipole moment of the trapped molecules. This could be understood as an additional contribution to the charge density due to charge transfer between graphene and trapped molecules. A substantial shift in $V_{Dirac}$ with the dipole moment of the trapped molecules has been observed (Fig.~\ref{Transport measurement}b). Such monotonic shift in $V_{Dirac}$ with dipole moment is attributed to the change of the effective gate capacitance due to underlying trapped molecules at the interface and should be considered as evidence for molecular trapping \cite{newaz2012probing,srivastava2015defect,xia2010effect}. For low dipole molecules, such as chlorobenzene, the interaction is relatively weaker, resulting in minimal charge transfer between graphene and these molecules. In contrast, trapped polar molecules interact more strongly with graphene. They tend to transfer more charge to graphene and create a charge cloud near the graphene basal plane leading to an increased bending height (inset of Fig.~\ref{Transport measurement}a). Dipole moment-dependent charge transfer between graphene and trapped molecules is experimentally demonstrated in our previous work \cite{khandelwal2023revealing}. This not only explains the shift in the $V_{Dirac}$, but also reveals the increase in the bending height with increase in the dipole moment of the trapped molecules.  
\begin{figure}
	\includegraphics[width=8.0 cm]{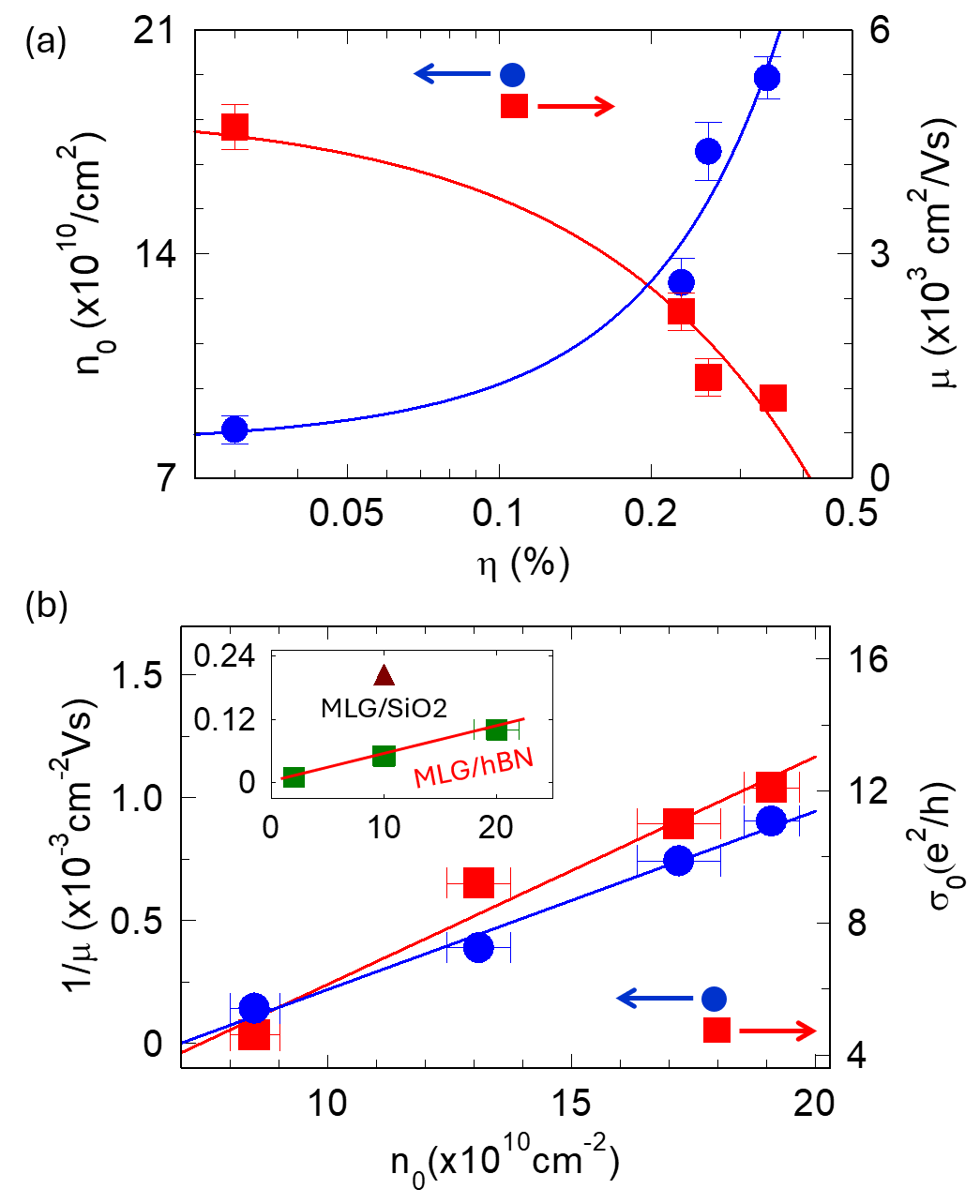}
	\caption{\textbf{Analysis of the transport measurements} (a) Residual carrier density, $n_0$ and mobility $\mu$ as a function of strain $\eta$. Symbols are the data points and solid lines are polynomial fit for guides of eyes. (b) Data in panel b plotted as $1/\mu$ as a function of $n_0$, showing linear relation. Panel b also shows experimental values of minimum conductivity $\sigma_0$ as a function of estimated $n_0$. Symbols are the data points and solid line is the linear fit.  The inset shows $1/\mu$ as a function of $n_0$ for graphene on $SiO_2$ and hBN as labeled (data in the inset are adapted from refs. \cite{couto2014random, ni2010resonant}). Data for $1/\mu$ vs $n_0$ on $SiO_2$ substrate is adapted with permission from ref. \cite{ni2010resonant}. Copyright {2010} American Chemical Society}.
	\label{Strain}
\end{figure}

We can interpret the transconductance of devices using the following relation based on the Drude model (equation \ref{eq2}) \cite{wang2020mobility,hong2009quantum,dean2010boron}, which relates device characteristics such as electron mobility and residual carrier density, that could potentially be affected by strain.
\begin{equation}
    G = \frac{1}{\frac{\alpha}{\mu e\sqrt{n_0^2+n^2}}+R}
    \label{eq2}
\end{equation}
where $\alpha$, $\mu$, $n_0$, and R are geometrical factor ($\alpha$ =1 in our case), density independent mobility, residual carrier density and serial resistance, respectively. The fitting results for $\mu$ and $n_0$ are plotted as a function of $\eta$ in Fig.~\ref{Strain}a. The $n_0$ shows a clear increase with increasing $\eta$, whereas $\mu$ decreases substantially. The $n_0$ increases from approximately $8.5\times10^{10}/cm^2$ to $\sim1.9\times10^{11}/cm^2$. Concurrently, $\mu$ decreases from 4500 $cm^2/Vs$ to 670 $cm^2/Vs$. Interestingly, when these parameters are plotted together ($1/\mu$ vs $n_0$), they show expected linear relation (Figure~\ref{Strain}b) \cite{gosling2021universal}. At low temperatures, $\mu$ is primarily limited by Coulomb and short-range scattering. In mono-layer graphene, mobility due to Coulomb scattering remains constant, while the mobility due to short-range scattering is inversely proportional to the carrier density \cite{srivastava2015defect}. This inverse relationship occurs because, for short-range scatterers, an increase in $n_0$ leads to more scattering events, thereby reducing the mobility of the carriers. To examine the exact origin of $n_0$, we monitored $1/\mu$ versus $n_0$ for pristine monolayer graphene placed on hBN and bare $SiO_2$ (Figure~\ref{Strain}b inset). We observe that at any fixed $n_0$ the $\frac{1}{\mu}$(hBN supported graphene) \textless $\frac{1}{\mu}$($SiO_2$ supported graphene) \textless  $\frac{1}{\mu}$(graphene with trapped molecules at the interface), indicating enhanced scattering rate when molecules are trapped at the interface and minimal contribution in case of hBN and bare $SiO_2$ substrates. This strain modulation with $n_0$ in our devices further corroborates that strain is influenced by electrostatic doping \cite{metten2016monitoring,chen2022charge}. Additionally, the linear variation of the experimentally observed $\sigma_0$ with the fitting parameter $n_0$ confirms that both parameters are affected by strain, which is controlled by the dipole moment of the trapped molecules. We would like to mention that molecular trapping at the interface can induce spatially varying electric fields through charge transfer, which may alter the density of states (DOS) and the band structure of graphene. We have made our argument to rule out this possibility as a potential threat to change the graphene’s transport behaviour, with a concise discussion provided in the SM, section S10 \cite{SM}.  

In summary, our study shows that molecular trapping at the graphene-substrate interface effectively controls strain in graphene devices using molecular dipole moments. Spectroscopic and transport measurements reveal a strong link between graphene strain and the dipole moment of trapped molecules, enabled by charge transfer. This method offers refined strain control graphene and could be extended to other van der Waals materials.\\

We thank WITec GmbH and AIRF, JNU for providing Raman facility. This research is
partially supported by SERB, Department of Science and
Technology, Government of India, through the project with
SERB Sanction Order No. CRG/2021/000776. 

\nocite{*}
\bibliography{bibliography}

\begin{thebibliography}{62}%
\makeatletter
\providecommand \@ifxundefined [1]{%
 \@ifx{#1\undefined}
}%
\providecommand \@ifnum [1]{%
 \ifnum #1\expandafter \@firstoftwo
 \else \expandafter \@secondoftwo
 \fi
}%
\providecommand \@ifx [1]{%
 \ifx #1\expandafter \@firstoftwo
 \else \expandafter \@secondoftwo
 \fi
}%
\providecommand \natexlab [1]{#1}%
\providecommand \enquote  [1]{``#1''}%
\providecommand \bibnamefont  [1]{#1}%
\providecommand \bibfnamefont [1]{#1}%
\providecommand \citenamefont [1]{#1}%
\providecommand \href@noop [0]{\@secondoftwo}%
\providecommand \href [0]{\begingroup \@sanitize@url \@href}%
\providecommand \@href[1]{\@@startlink{#1}\@@href}%
\providecommand \@@href[1]{\endgroup#1\@@endlink}%
\providecommand \@sanitize@url [0]{\catcode `\\12\catcode `\$12\catcode `\&12\catcode `\#12\catcode `\^12\catcode `\_12\catcode `\%12\relax}%
\providecommand \@@startlink[1]{}%
\providecommand \@@endlink[0]{}%
\providecommand \url  [0]{\begingroup\@sanitize@url \@url }%
\providecommand \@url [1]{\endgroup\@href {#1}{\urlprefix }}%
\providecommand \urlprefix  [0]{URL }%
\providecommand \Eprint [0]{\href }%
\providecommand \doibase [0]{https://doi.org/}%
\providecommand \selectlanguage [0]{\@gobble}%
\providecommand \bibinfo  [0]{\@secondoftwo}%
\providecommand \bibfield  [0]{\@secondoftwo}%
\providecommand \translation [1]{[#1]}%
\providecommand \BibitemOpen [0]{}%
\providecommand \bibitemStop [0]{}%
\providecommand \bibitemNoStop [0]{.\EOS\space}%
\providecommand \EOS [0]{\spacefactor3000\relax}%
\providecommand \BibitemShut  [1]{\csname bibitem#1\endcsname}%
\let\auto@bib@innerbib\@empty
\bibitem [{\citenamefont {Ghosh}\ \emph {et~al.}(2023)\citenamefont {Ghosh}, \citenamefont {Zheng}, \citenamefont {Radhakrishnan}, \citenamefont {Schranghamer},\ and\ \citenamefont {Das}}]{ghosh2023graphene}%
  \BibitemOpen
  \bibfield  {author} {\bibinfo {author} {\bibfnamefont {S.}~\bibnamefont {Ghosh}}, \bibinfo {author} {\bibfnamefont {Y.}~\bibnamefont {Zheng}}, \bibinfo {author} {\bibfnamefont {S.~S.}\ \bibnamefont {Radhakrishnan}}, \bibinfo {author} {\bibfnamefont {T.~F.}\ \bibnamefont {Schranghamer}},\ and\ \bibinfo {author} {\bibfnamefont {S.}~\bibnamefont {Das}},\ }\bibfield  {title} {\bibinfo {title} {A graphene-based straintronic physically unclonable function},\ }\href {https://pubs.acs.org/doi/10.1021/acs.nanolett.3c01145} {\bibfield  {journal} {\bibinfo  {journal} {Nano letters}\ }\textbf {\bibinfo {volume} {23}},\ \bibinfo {pages} {5171} (\bibinfo {year} {2023})}\BibitemShut {NoStop}%
\bibitem [{\citenamefont {Hou}\ \emph {et~al.}(2024)\citenamefont {Hou}, \citenamefont {Zhou}, \citenamefont {Xue}, \citenamefont {Yu}, \citenamefont {Han}, \citenamefont {Zhang},\ and\ \citenamefont {Lu}}]{hou2024strain}%
  \BibitemOpen
  \bibfield  {author} {\bibinfo {author} {\bibfnamefont {Y.}~\bibnamefont {Hou}}, \bibinfo {author} {\bibfnamefont {J.}~\bibnamefont {Zhou}}, \bibinfo {author} {\bibfnamefont {M.}~\bibnamefont {Xue}}, \bibinfo {author} {\bibfnamefont {M.}~\bibnamefont {Yu}}, \bibinfo {author} {\bibfnamefont {Y.}~\bibnamefont {Han}}, \bibinfo {author} {\bibfnamefont {Z.}~\bibnamefont {Zhang}},\ and\ \bibinfo {author} {\bibfnamefont {Y.}~\bibnamefont {Lu}},\ }\bibfield  {title} {\bibinfo {title} {Strain engineering of twisted bilayer graphene: The rise of strain-twistronics},\ }\href {https://onlinelibrary.wiley.com/doi/full/10.1002/smll.202311185} {\bibfield  {journal} {\bibinfo  {journal} {Small}\ ,\ \bibinfo {pages} {2311185}} (\bibinfo {year} {2024})}\BibitemShut {NoStop}%
\bibitem [{\citenamefont {Solomenko}\ \emph {et~al.}(2023)\citenamefont {Solomenko}, \citenamefont {Sahalianov}, \citenamefont {Radchenko},\ and\ \citenamefont {Tatarenko}}]{solomenko2023straintronics}%
  \BibitemOpen
  \bibfield  {author} {\bibinfo {author} {\bibfnamefont {A.~G.}\ \bibnamefont {Solomenko}}, \bibinfo {author} {\bibfnamefont {I.~Y.}\ \bibnamefont {Sahalianov}}, \bibinfo {author} {\bibfnamefont {T.~M.}\ \bibnamefont {Radchenko}},\ and\ \bibinfo {author} {\bibfnamefont {V.~A.}\ \bibnamefont {Tatarenko}},\ }\bibfield  {title} {\bibinfo {title} {Straintronics in phosphorene via tensile vs shear strains and their combinations for manipulating the band gap},\ }\href {https://www.nature.com/articles/s41598-023-40541-7} {\bibfield  {journal} {\bibinfo  {journal} {Scientific Reports}\ }\textbf {\bibinfo {volume} {13}},\ \bibinfo {pages} {13444} (\bibinfo {year} {2023})}\BibitemShut {NoStop}%
\bibitem [{\citenamefont {Peng}\ \emph {et~al.}(2020)\citenamefont {Peng}, \citenamefont {Chen}, \citenamefont {Fan}, \citenamefont {Srolovitz},\ and\ \citenamefont {Lei}}]{peng2020strain}%
  \BibitemOpen
  \bibfield  {author} {\bibinfo {author} {\bibfnamefont {Z.}~\bibnamefont {Peng}}, \bibinfo {author} {\bibfnamefont {X.}~\bibnamefont {Chen}}, \bibinfo {author} {\bibfnamefont {Y.}~\bibnamefont {Fan}}, \bibinfo {author} {\bibfnamefont {D.~J.}\ \bibnamefont {Srolovitz}},\ and\ \bibinfo {author} {\bibfnamefont {D.}~\bibnamefont {Lei}},\ }\bibfield  {title} {\bibinfo {title} {Strain engineering of 2d semiconductors and graphene: from strain fields to band-structure tuning and photonic applications},\ }\href {https://www.nature.com/articles/s41377-020-00421-5} {\bibfield  {journal} {\bibinfo  {journal} {Light: Science \& Applications}\ }\textbf {\bibinfo {volume} {9}},\ \bibinfo {pages} {190} (\bibinfo {year} {2020})}\BibitemShut {NoStop}%
\bibitem [{\citenamefont {Liu}\ \emph {et~al.}(2014)\citenamefont {Liu}, \citenamefont {Amani}, \citenamefont {Najmaei}, \citenamefont {Xu}, \citenamefont {Zou}, \citenamefont {Zhou}, \citenamefont {Yu}, \citenamefont {Qiu}, \citenamefont {Birdwell}, \citenamefont {Crowne} \emph {et~al.}}]{liu2014strain}%
  \BibitemOpen
  \bibfield  {author} {\bibinfo {author} {\bibfnamefont {Z.}~\bibnamefont {Liu}}, \bibinfo {author} {\bibfnamefont {M.}~\bibnamefont {Amani}}, \bibinfo {author} {\bibfnamefont {S.}~\bibnamefont {Najmaei}}, \bibinfo {author} {\bibfnamefont {Q.}~\bibnamefont {Xu}}, \bibinfo {author} {\bibfnamefont {X.}~\bibnamefont {Zou}}, \bibinfo {author} {\bibfnamefont {W.}~\bibnamefont {Zhou}}, \bibinfo {author} {\bibfnamefont {T.}~\bibnamefont {Yu}}, \bibinfo {author} {\bibfnamefont {C.}~\bibnamefont {Qiu}}, \bibinfo {author} {\bibfnamefont {A.~G.}\ \bibnamefont {Birdwell}}, \bibinfo {author} {\bibfnamefont {F.~J.}\ \bibnamefont {Crowne}}, \emph {et~al.},\ }\bibfield  {title} {\bibinfo {title} {Strain and structure heterogeneity in mos2 atomic layers grown by chemical vapour deposition},\ }\href {https://www.nature.com/articles/ncomms6246} {\bibfield  {journal} {\bibinfo  {journal} {Nature communications}\ }\textbf {\bibinfo {volume} {5}},\ \bibinfo {pages} {5246} (\bibinfo {year} {2014})}\BibitemShut {NoStop}%
\bibitem [{\citenamefont {Mennel}\ \emph {et~al.}(2018)\citenamefont {Mennel}, \citenamefont {Furchi}, \citenamefont {Wachter}, \citenamefont {Paur}, \citenamefont {Polyushkin},\ and\ \citenamefont {Mueller}}]{mennel2018optical}%
  \BibitemOpen
  \bibfield  {author} {\bibinfo {author} {\bibfnamefont {L.}~\bibnamefont {Mennel}}, \bibinfo {author} {\bibfnamefont {M.~M.}\ \bibnamefont {Furchi}}, \bibinfo {author} {\bibfnamefont {S.}~\bibnamefont {Wachter}}, \bibinfo {author} {\bibfnamefont {M.}~\bibnamefont {Paur}}, \bibinfo {author} {\bibfnamefont {D.~K.}\ \bibnamefont {Polyushkin}},\ and\ \bibinfo {author} {\bibfnamefont {T.}~\bibnamefont {Mueller}},\ }\bibfield  {title} {\bibinfo {title} {Optical imaging of strain in two-dimensional crystals},\ }\href {https://www.nature.com/articles/s41467-018-02830-y} {\bibfield  {journal} {\bibinfo  {journal} {Nature communications}\ }\textbf {\bibinfo {volume} {9}},\ \bibinfo {pages} {516} (\bibinfo {year} {2018})}\BibitemShut {NoStop}%
\bibitem [{\citenamefont {Yang}\ \emph {et~al.}(2021)\citenamefont {Yang}, \citenamefont {Chen},\ and\ \citenamefont {Jiang}}]{yang2021strain}%
  \BibitemOpen
  \bibfield  {author} {\bibinfo {author} {\bibfnamefont {S.}~\bibnamefont {Yang}}, \bibinfo {author} {\bibfnamefont {Y.}~\bibnamefont {Chen}},\ and\ \bibinfo {author} {\bibfnamefont {C.}~\bibnamefont {Jiang}},\ }\bibfield  {title} {\bibinfo {title} {Strain engineering of two-dimensional materials: Methods, properties, and applications},\ }\href {https://onlinelibrary.wiley.com/doi/full/10.1002/inf2.12177} {\bibfield  {journal} {\bibinfo  {journal} {InfoMat}\ }\textbf {\bibinfo {volume} {3}},\ \bibinfo {pages} {397} (\bibinfo {year} {2021})}\BibitemShut {NoStop}%
\bibitem [{\citenamefont {Lee}\ \emph {et~al.}(2012)\citenamefont {Lee}, \citenamefont {Ahn}, \citenamefont {Shim}, \citenamefont {Lee},\ and\ \citenamefont {Ryu}}]{lee2012optical}%
  \BibitemOpen
  \bibfield  {author} {\bibinfo {author} {\bibfnamefont {J.~E.}\ \bibnamefont {Lee}}, \bibinfo {author} {\bibfnamefont {G.}~\bibnamefont {Ahn}}, \bibinfo {author} {\bibfnamefont {J.}~\bibnamefont {Shim}}, \bibinfo {author} {\bibfnamefont {Y.~S.}\ \bibnamefont {Lee}},\ and\ \bibinfo {author} {\bibfnamefont {S.}~\bibnamefont {Ryu}},\ }\bibfield  {title} {\bibinfo {title} {Optical separation of mechanical strain from charge doping in graphene},\ }\href {https://www.nature.com/articles/ncomms2022} {\bibfield  {journal} {\bibinfo  {journal} {Nature communications}\ }\textbf {\bibinfo {volume} {3}},\ \bibinfo {pages} {1024} (\bibinfo {year} {2012})}\BibitemShut {NoStop}%
\bibitem [{\citenamefont {Lee}\ \emph {et~al.}(2008)\citenamefont {Lee}, \citenamefont {Wei}, \citenamefont {Kysar},\ and\ \citenamefont {Hone}}]{lee2008measurement}%
  \BibitemOpen
  \bibfield  {author} {\bibinfo {author} {\bibfnamefont {C.}~\bibnamefont {Lee}}, \bibinfo {author} {\bibfnamefont {X.}~\bibnamefont {Wei}}, \bibinfo {author} {\bibfnamefont {J.~W.}\ \bibnamefont {Kysar}},\ and\ \bibinfo {author} {\bibfnamefont {J.}~\bibnamefont {Hone}},\ }\bibfield  {title} {\bibinfo {title} {Measurement of the elastic properties and intrinsic strength of monolayer graphene},\ }\href {http://10.1126/science.1157996} {\bibfield  {journal} {\bibinfo  {journal} {science}\ }\textbf {\bibinfo {volume} {321}},\ \bibinfo {pages} {385} (\bibinfo {year} {2008})}\BibitemShut {NoStop}%
\bibitem [{\citenamefont {Choi}\ \emph {et~al.}(2010)\citenamefont {Choi}, \citenamefont {Jhi},\ and\ \citenamefont {Son}}]{choi2010effects}%
  \BibitemOpen
  \bibfield  {author} {\bibinfo {author} {\bibfnamefont {S.-M.}\ \bibnamefont {Choi}}, \bibinfo {author} {\bibfnamefont {S.-H.}\ \bibnamefont {Jhi}},\ and\ \bibinfo {author} {\bibfnamefont {Y.-W.}\ \bibnamefont {Son}},\ }\bibfield  {title} {\bibinfo {title} {Effects of strain on electronic properties of graphene},\ }\href {https://journals.aps.org/prb/abstract/10.1103/PhysRevB.81.081407} {\bibfield  {journal} {\bibinfo  {journal} {Physical Review B—Condensed Matter and Materials Physics}\ }\textbf {\bibinfo {volume} {81}},\ \bibinfo {pages} {081407} (\bibinfo {year} {2010})}\BibitemShut {NoStop}%
\bibitem [{\citenamefont {Postorino}\ \emph {et~al.}(2020)\citenamefont {Postorino}, \citenamefont {Grassano}, \citenamefont {D’Alessandro}, \citenamefont {Pianetti}, \citenamefont {Pulci},\ and\ \citenamefont {Palummo}}]{postorino2020strain}%
  \BibitemOpen
  \bibfield  {author} {\bibinfo {author} {\bibfnamefont {S.}~\bibnamefont {Postorino}}, \bibinfo {author} {\bibfnamefont {D.}~\bibnamefont {Grassano}}, \bibinfo {author} {\bibfnamefont {M.}~\bibnamefont {D’Alessandro}}, \bibinfo {author} {\bibfnamefont {A.}~\bibnamefont {Pianetti}}, \bibinfo {author} {\bibfnamefont {O.}~\bibnamefont {Pulci}},\ and\ \bibinfo {author} {\bibfnamefont {M.}~\bibnamefont {Palummo}},\ }\bibfield  {title} {\bibinfo {title} {Strain-induced effects on the electronic properties of 2d materials},\ }\href {https://journals.sagepub.com/doi/full/10.1177/1847980420902569} {\bibfield  {journal} {\bibinfo  {journal} {Nanomaterials and Nanotechnology}\ }\textbf {\bibinfo {volume} {10}},\ \bibinfo {pages} {1847980420902569} (\bibinfo {year} {2020})}\BibitemShut {NoStop}%
\bibitem [{\citenamefont {Parker}\ \emph {et~al.}(2021)\citenamefont {Parker}, \citenamefont {Soejima}, \citenamefont {Hauschild}, \citenamefont {Zaletel},\ and\ \citenamefont {Bultinck}}]{parker2021strain}%
  \BibitemOpen
  \bibfield  {author} {\bibinfo {author} {\bibfnamefont {D.~E.}\ \bibnamefont {Parker}}, \bibinfo {author} {\bibfnamefont {T.}~\bibnamefont {Soejima}}, \bibinfo {author} {\bibfnamefont {J.}~\bibnamefont {Hauschild}}, \bibinfo {author} {\bibfnamefont {M.~P.}\ \bibnamefont {Zaletel}},\ and\ \bibinfo {author} {\bibfnamefont {N.}~\bibnamefont {Bultinck}},\ }\bibfield  {title} {\bibinfo {title} {Strain-induced quantum phase transitions in magic-angle graphene},\ }\href {https://journals.aps.org/prl/abstract/10.1103/PhysRevLett.127.027601} {\bibfield  {journal} {\bibinfo  {journal} {Physical review letters}\ }\textbf {\bibinfo {volume} {127}},\ \bibinfo {pages} {027601} (\bibinfo {year} {2021})}\BibitemShut {NoStop}%
\bibitem [{\citenamefont {Settnes}\ \emph {et~al.}(2016)\citenamefont {Settnes}, \citenamefont {Power},\ and\ \citenamefont {Jauho}}]{settnes2016pseudomagnetic}%
  \BibitemOpen
  \bibfield  {author} {\bibinfo {author} {\bibfnamefont {M.}~\bibnamefont {Settnes}}, \bibinfo {author} {\bibfnamefont {S.~R.}\ \bibnamefont {Power}},\ and\ \bibinfo {author} {\bibfnamefont {A.-P.}\ \bibnamefont {Jauho}},\ }\bibfield  {title} {\bibinfo {title} {Pseudomagnetic fields and triaxial strain in graphene},\ }\href {https://journals.aps.org/prb/abstract/10.1103/PhysRevB.93.035456} {\bibfield  {journal} {\bibinfo  {journal} {Physical Review B}\ }\textbf {\bibinfo {volume} {93}},\ \bibinfo {pages} {035456} (\bibinfo {year} {2016})}\BibitemShut {NoStop}%
\bibitem [{\citenamefont {Levy}\ \emph {et~al.}(2010)\citenamefont {Levy}, \citenamefont {Burke}, \citenamefont {Meaker}, \citenamefont {Panlasigui}, \citenamefont {Zettl}, \citenamefont {Guinea}, \citenamefont {Neto},\ and\ \citenamefont {Crommie}}]{levy2010strain}%
  \BibitemOpen
  \bibfield  {author} {\bibinfo {author} {\bibfnamefont {N.}~\bibnamefont {Levy}}, \bibinfo {author} {\bibfnamefont {S.}~\bibnamefont {Burke}}, \bibinfo {author} {\bibfnamefont {K.}~\bibnamefont {Meaker}}, \bibinfo {author} {\bibfnamefont {M.}~\bibnamefont {Panlasigui}}, \bibinfo {author} {\bibfnamefont {A.}~\bibnamefont {Zettl}}, \bibinfo {author} {\bibfnamefont {F.}~\bibnamefont {Guinea}}, \bibinfo {author} {\bibfnamefont {A.~C.}\ \bibnamefont {Neto}},\ and\ \bibinfo {author} {\bibfnamefont {M.~F.}\ \bibnamefont {Crommie}},\ }\bibfield  {title} {\bibinfo {title} {Strain-induced pseudo--magnetic fields greater than 300 tesla in graphene nanobubbles},\ }\href {http://10.1126/science.1191700} {\bibfield  {journal} {\bibinfo  {journal} {Science}\ }\textbf {\bibinfo {volume} {329}},\ \bibinfo {pages} {544} (\bibinfo {year} {2010})}\BibitemShut {NoStop}%
\bibitem [{\citenamefont {Zhang}\ \emph {et~al.}(2022)\citenamefont {Zhang}, \citenamefont {Jin}, \citenamefont {Liu}, \citenamefont {Ren}, \citenamefont {Chen}, \citenamefont {Zhao},\ and\ \citenamefont {Zhao}}]{zhang2022strain}%
  \BibitemOpen
  \bibfield  {author} {\bibinfo {author} {\bibfnamefont {Y.}~\bibnamefont {Zhang}}, \bibinfo {author} {\bibfnamefont {Y.}~\bibnamefont {Jin}}, \bibinfo {author} {\bibfnamefont {J.}~\bibnamefont {Liu}}, \bibinfo {author} {\bibfnamefont {Q.}~\bibnamefont {Ren}}, \bibinfo {author} {\bibfnamefont {Z.}~\bibnamefont {Chen}}, \bibinfo {author} {\bibfnamefont {Y.}~\bibnamefont {Zhao}},\ and\ \bibinfo {author} {\bibfnamefont {P.}~\bibnamefont {Zhao}},\ }\bibfield  {title} {\bibinfo {title} {Strain engineering of graphene on rigid substrates},\ }\href {https://www.ncbi.nlm.nih.gov/pmc/articles/PMC9680924/} {\bibfield  {journal} {\bibinfo  {journal} {Nanoscale Advances}\ }\textbf {\bibinfo {volume} {4}},\ \bibinfo {pages} {5056} (\bibinfo {year} {2022})}\BibitemShut {NoStop}%
\bibitem [{\citenamefont {Banerjee}\ \emph {et~al.}(2020)\citenamefont {Banerjee}, \citenamefont {Nguyen}, \citenamefont {Granzier-Nakajima}, \citenamefont {Pabbi}, \citenamefont {Lherbier}, \citenamefont {Binion}, \citenamefont {Charlier}, \citenamefont {Terrones},\ and\ \citenamefont {Hudson}}]{banerjee2020strain}%
  \BibitemOpen
  \bibfield  {author} {\bibinfo {author} {\bibfnamefont {R.}~\bibnamefont {Banerjee}}, \bibinfo {author} {\bibfnamefont {V.-H.}\ \bibnamefont {Nguyen}}, \bibinfo {author} {\bibfnamefont {T.}~\bibnamefont {Granzier-Nakajima}}, \bibinfo {author} {\bibfnamefont {L.}~\bibnamefont {Pabbi}}, \bibinfo {author} {\bibfnamefont {A.}~\bibnamefont {Lherbier}}, \bibinfo {author} {\bibfnamefont {A.~R.}\ \bibnamefont {Binion}}, \bibinfo {author} {\bibfnamefont {J.-C.}\ \bibnamefont {Charlier}}, \bibinfo {author} {\bibfnamefont {M.}~\bibnamefont {Terrones}},\ and\ \bibinfo {author} {\bibfnamefont {E.~W.}\ \bibnamefont {Hudson}},\ }\bibfield  {title} {\bibinfo {title} {Strain modulated superlattices in graphene},\ }\href {https://pubs.acs.org/doi/10.1021/acs.nanolett.9b05108?goto=articleMetrics&ref=pdf} {\bibfield  {journal} {\bibinfo  {journal} {Nano letters}\ }\textbf {\bibinfo {volume} {20}},\ \bibinfo {pages} {3113} (\bibinfo {year} {2020})}\BibitemShut {NoStop}%
\bibitem [{\citenamefont {Locatelli}\ \emph {et~al.}(2010)\citenamefont {Locatelli}, \citenamefont {Knox}, \citenamefont {Cvetko}, \citenamefont {Mentes}, \citenamefont {Nino}, \citenamefont {Wang}, \citenamefont {Yilmaz}, \citenamefont {Kim}, \citenamefont {Osgood~Jr},\ and\ \citenamefont {Morgante}}]{locatelli2010corrugation}%
  \BibitemOpen
  \bibfield  {author} {\bibinfo {author} {\bibfnamefont {A.}~\bibnamefont {Locatelli}}, \bibinfo {author} {\bibfnamefont {K.~R.}\ \bibnamefont {Knox}}, \bibinfo {author} {\bibfnamefont {D.}~\bibnamefont {Cvetko}}, \bibinfo {author} {\bibfnamefont {T.~O.}\ \bibnamefont {Mentes}}, \bibinfo {author} {\bibfnamefont {M.~A.}\ \bibnamefont {Nino}}, \bibinfo {author} {\bibfnamefont {S.}~\bibnamefont {Wang}}, \bibinfo {author} {\bibfnamefont {M.~B.}\ \bibnamefont {Yilmaz}}, \bibinfo {author} {\bibfnamefont {P.}~\bibnamefont {Kim}}, \bibinfo {author} {\bibfnamefont {R.~M.}\ \bibnamefont {Osgood~Jr}},\ and\ \bibinfo {author} {\bibfnamefont {A.}~\bibnamefont {Morgante}},\ }\bibfield  {title} {\bibinfo {title} {Corrugation in exfoliated graphene: an electron microscopy and diffraction study},\ }\href {https://doi.org/10.1021/nn101116n} {\bibfield  {journal} {\bibinfo  {journal} {ACS nano}\ }\textbf {\bibinfo {volume} {4}},\ \bibinfo {pages} {4879} (\bibinfo {year} {2010})}\BibitemShut {NoStop}%
\bibitem [{\citenamefont {Zihlmann}\ \emph {et~al.}(2020)\citenamefont {Zihlmann}, \citenamefont {Makk}, \citenamefont {Rehmann}, \citenamefont {Wang}, \citenamefont {Kedves}, \citenamefont {Indolese}, \citenamefont {Watanabe}, \citenamefont {Taniguchi}, \citenamefont {Zumb{\"u}hl},\ and\ \citenamefont {Sch{\"o}nenberger}}]{zihlmann2020out}%
  \BibitemOpen
  \bibfield  {author} {\bibinfo {author} {\bibfnamefont {S.}~\bibnamefont {Zihlmann}}, \bibinfo {author} {\bibfnamefont {P.}~\bibnamefont {Makk}}, \bibinfo {author} {\bibfnamefont {M.~K.}\ \bibnamefont {Rehmann}}, \bibinfo {author} {\bibfnamefont {L.}~\bibnamefont {Wang}}, \bibinfo {author} {\bibfnamefont {M.}~\bibnamefont {Kedves}}, \bibinfo {author} {\bibfnamefont {D.~I.}\ \bibnamefont {Indolese}}, \bibinfo {author} {\bibfnamefont {K.}~\bibnamefont {Watanabe}}, \bibinfo {author} {\bibfnamefont {T.}~\bibnamefont {Taniguchi}}, \bibinfo {author} {\bibfnamefont {D.~M.}\ \bibnamefont {Zumb{\"u}hl}},\ and\ \bibinfo {author} {\bibfnamefont {C.}~\bibnamefont {Sch{\"o}nenberger}},\ }\bibfield  {title} {\bibinfo {title} {Out-of-plane corrugations in graphene based van der waals heterostructures},\ }\href {https://journals.aps.org/prb/abstract/10.1103/PhysRevB.102.195404} {\bibfield  {journal} {\bibinfo  {journal} {Physical Review B}\ }\textbf {\bibinfo {volume} {102}},\ \bibinfo {pages} {195404} (\bibinfo {year}
  {2020})}\BibitemShut {NoStop}%
\bibitem [{\citenamefont {Couto}\ \emph {et~al.}(2014)\citenamefont {Couto}, \citenamefont {Costanzo}, \citenamefont {Engels}, \citenamefont {Ki}, \citenamefont {Watanabe}, \citenamefont {Taniguchi}, \citenamefont {Stampfer}, \citenamefont {Guinea},\ and\ \citenamefont {Morpurgo}}]{couto2014random}%
  \BibitemOpen
  \bibfield  {author} {\bibinfo {author} {\bibfnamefont {N.~J.}\ \bibnamefont {Couto}}, \bibinfo {author} {\bibfnamefont {D.}~\bibnamefont {Costanzo}}, \bibinfo {author} {\bibfnamefont {S.}~\bibnamefont {Engels}}, \bibinfo {author} {\bibfnamefont {D.-K.}\ \bibnamefont {Ki}}, \bibinfo {author} {\bibfnamefont {K.}~\bibnamefont {Watanabe}}, \bibinfo {author} {\bibfnamefont {T.}~\bibnamefont {Taniguchi}}, \bibinfo {author} {\bibfnamefont {C.}~\bibnamefont {Stampfer}}, \bibinfo {author} {\bibfnamefont {F.}~\bibnamefont {Guinea}},\ and\ \bibinfo {author} {\bibfnamefont {A.~F.}\ \bibnamefont {Morpurgo}},\ }\bibfield  {title} {\bibinfo {title} {Random strain fluctuations as dominant disorder source for high-quality on-substrate graphene devices},\ }\href {https://journals.aps.org/prx/abstract/10.1103/PhysRevX.4.041019} {\bibfield  {journal} {\bibinfo  {journal} {Physical Review X}\ }\textbf {\bibinfo {volume} {4}},\ \bibinfo {pages} {041019} (\bibinfo {year} {2014})}\BibitemShut {NoStop}%
\bibitem [{\citenamefont {Wang}\ \emph {et~al.}(2020)\citenamefont {Wang}, \citenamefont {Makk}, \citenamefont {Zihlmann}, \citenamefont {Baumgartner}, \citenamefont {Indolese}, \citenamefont {Watanabe}, \citenamefont {Taniguchi},\ and\ \citenamefont {Sch{\"o}nenberger}}]{wang2020mobility}%
  \BibitemOpen
  \bibfield  {author} {\bibinfo {author} {\bibfnamefont {L.}~\bibnamefont {Wang}}, \bibinfo {author} {\bibfnamefont {P.}~\bibnamefont {Makk}}, \bibinfo {author} {\bibfnamefont {S.}~\bibnamefont {Zihlmann}}, \bibinfo {author} {\bibfnamefont {A.}~\bibnamefont {Baumgartner}}, \bibinfo {author} {\bibfnamefont {D.~I.}\ \bibnamefont {Indolese}}, \bibinfo {author} {\bibfnamefont {K.}~\bibnamefont {Watanabe}}, \bibinfo {author} {\bibfnamefont {T.}~\bibnamefont {Taniguchi}},\ and\ \bibinfo {author} {\bibfnamefont {C.}~\bibnamefont {Sch{\"o}nenberger}},\ }\bibfield  {title} {\bibinfo {title} {Mobility enhancement in graphene by in situ reduction of random strain fluctuations},\ }\href {https://journals.aps.org/prl/abstract/10.1103/PhysRevLett.124.157701} {\bibfield  {journal} {\bibinfo  {journal} {Physical review letters}\ }\textbf {\bibinfo {volume} {124}},\ \bibinfo {pages} {157701} (\bibinfo {year} {2020})}\BibitemShut {NoStop}%
\bibitem [{\citenamefont {Ni}\ \emph {et~al.}(2008)\citenamefont {Ni}, \citenamefont {Yu}, \citenamefont {Lu}, \citenamefont {Wang}, \citenamefont {Feng},\ and\ \citenamefont {Shen}}]{ni2008uniaxial}%
  \BibitemOpen
  \bibfield  {author} {\bibinfo {author} {\bibfnamefont {Z.~H.}\ \bibnamefont {Ni}}, \bibinfo {author} {\bibfnamefont {T.}~\bibnamefont {Yu}}, \bibinfo {author} {\bibfnamefont {Y.~H.}\ \bibnamefont {Lu}}, \bibinfo {author} {\bibfnamefont {Y.~Y.}\ \bibnamefont {Wang}}, \bibinfo {author} {\bibfnamefont {Y.~P.}\ \bibnamefont {Feng}},\ and\ \bibinfo {author} {\bibfnamefont {Z.~X.}\ \bibnamefont {Shen}},\ }\bibfield  {title} {\bibinfo {title} {Uniaxial strain on graphene: Raman spectroscopy study and band-gap opening},\ }\href {https://pubs.acs.org/doi/10.1021/nn800459e} {\bibfield  {journal} {\bibinfo  {journal} {ACS nano}\ }\textbf {\bibinfo {volume} {2}},\ \bibinfo {pages} {2301} (\bibinfo {year} {2008})}\BibitemShut {NoStop}%
\bibitem [{\citenamefont {Rold{\'a}n}\ \emph {et~al.}(2015)\citenamefont {Rold{\'a}n}, \citenamefont {Castellanos-Gomez}, \citenamefont {Cappelluti},\ and\ \citenamefont {Guinea}}]{roldan2015strain}%
  \BibitemOpen
  \bibfield  {author} {\bibinfo {author} {\bibfnamefont {R.}~\bibnamefont {Rold{\'a}n}}, \bibinfo {author} {\bibfnamefont {A.}~\bibnamefont {Castellanos-Gomez}}, \bibinfo {author} {\bibfnamefont {E.}~\bibnamefont {Cappelluti}},\ and\ \bibinfo {author} {\bibfnamefont {F.}~\bibnamefont {Guinea}},\ }\bibfield  {title} {\bibinfo {title} {Strain engineering in semiconducting two-dimensional crystals},\ }\href {https://iopscience.iop.org/article/10.1088/0953-8984/27/31/313201/meta} {\bibfield  {journal} {\bibinfo  {journal} {Journal of Physics: Condensed Matter}\ }\textbf {\bibinfo {volume} {27}},\ \bibinfo {pages} {313201} (\bibinfo {year} {2015})}\BibitemShut {NoStop}%
\bibitem [{\citenamefont {Mohiuddin}\ \emph {et~al.}(2009)\citenamefont {Mohiuddin}, \citenamefont {Lombardo}, \citenamefont {Nair}, \citenamefont {Bonetti}, \citenamefont {Savini}, \citenamefont {Jalil}, \citenamefont {Bonini}, \citenamefont {Basko}, \citenamefont {Galiotis}, \citenamefont {Marzari} \emph {et~al.}}]{mohiuddin2009uniaxial}%
  \BibitemOpen
  \bibfield  {author} {\bibinfo {author} {\bibfnamefont {T.}~\bibnamefont {Mohiuddin}}, \bibinfo {author} {\bibfnamefont {A.}~\bibnamefont {Lombardo}}, \bibinfo {author} {\bibfnamefont {R.}~\bibnamefont {Nair}}, \bibinfo {author} {\bibfnamefont {A.}~\bibnamefont {Bonetti}}, \bibinfo {author} {\bibfnamefont {G.}~\bibnamefont {Savini}}, \bibinfo {author} {\bibfnamefont {R.}~\bibnamefont {Jalil}}, \bibinfo {author} {\bibfnamefont {N.}~\bibnamefont {Bonini}}, \bibinfo {author} {\bibfnamefont {D.}~\bibnamefont {Basko}}, \bibinfo {author} {\bibfnamefont {C.}~\bibnamefont {Galiotis}}, \bibinfo {author} {\bibfnamefont {N.}~\bibnamefont {Marzari}}, \emph {et~al.},\ }\bibfield  {title} {\bibinfo {title} {Uniaxial strain in graphene by raman spectroscopy: G peak splitting, gr{\"u}neisen parameters, and sample orientation},\ }\href {https://journals.aps.org/prb/abstract/10.1103/PhysRevB.79.205433} {\bibfield  {journal} {\bibinfo  {journal} {Physical Review B—Condensed Matter and Materials Physics}\ }\textbf {\bibinfo
  {volume} {79}},\ \bibinfo {pages} {205433} (\bibinfo {year} {2009})}\BibitemShut {NoStop}%
\bibitem [{\citenamefont {Won}\ \emph {et~al.}(2022)\citenamefont {Won}, \citenamefont {Lee}, \citenamefont {Jung}, \citenamefont {Kwon}, \citenamefont {Gebredingle}, \citenamefont {Lim}, \citenamefont {Kim}, \citenamefont {Jeong},\ and\ \citenamefont {Lee}}]{won2022raman}%
  \BibitemOpen
  \bibfield  {author} {\bibinfo {author} {\bibfnamefont {K.}~\bibnamefont {Won}}, \bibinfo {author} {\bibfnamefont {C.}~\bibnamefont {Lee}}, \bibinfo {author} {\bibfnamefont {J.}~\bibnamefont {Jung}}, \bibinfo {author} {\bibfnamefont {S.}~\bibnamefont {Kwon}}, \bibinfo {author} {\bibfnamefont {Y.}~\bibnamefont {Gebredingle}}, \bibinfo {author} {\bibfnamefont {J.~G.}\ \bibnamefont {Lim}}, \bibinfo {author} {\bibfnamefont {M.~K.}\ \bibnamefont {Kim}}, \bibinfo {author} {\bibfnamefont {M.~S.}\ \bibnamefont {Jeong}},\ and\ \bibinfo {author} {\bibfnamefont {C.}~\bibnamefont {Lee}},\ }\bibfield  {title} {\bibinfo {title} {Raman scattering measurement of suspended graphene under extreme strain induced by nanoindentation},\ }\href {https://onlinelibrary.wiley.com/doi/abs/10.1002/adma.202200946} {\bibfield  {journal} {\bibinfo  {journal} {Advanced Materials}\ }\textbf {\bibinfo {volume} {34}},\ \bibinfo {pages} {2200946} (\bibinfo {year} {2022})}\BibitemShut {NoStop}%
\bibitem [{\citenamefont {Ahn}\ \emph {et~al.}(2017)\citenamefont {Ahn}, \citenamefont {Amani}, \citenamefont {Rasool}, \citenamefont {Lien}, \citenamefont {Mastandrea}, \citenamefont {Ager~III}, \citenamefont {Dubey}, \citenamefont {Chrzan}, \citenamefont {Minor},\ and\ \citenamefont {Javey}}]{ahn2017strain}%
  \BibitemOpen
  \bibfield  {author} {\bibinfo {author} {\bibfnamefont {G.~H.}\ \bibnamefont {Ahn}}, \bibinfo {author} {\bibfnamefont {M.}~\bibnamefont {Amani}}, \bibinfo {author} {\bibfnamefont {H.}~\bibnamefont {Rasool}}, \bibinfo {author} {\bibfnamefont {D.-H.}\ \bibnamefont {Lien}}, \bibinfo {author} {\bibfnamefont {J.~P.}\ \bibnamefont {Mastandrea}}, \bibinfo {author} {\bibfnamefont {J.~W.}\ \bibnamefont {Ager~III}}, \bibinfo {author} {\bibfnamefont {M.}~\bibnamefont {Dubey}}, \bibinfo {author} {\bibfnamefont {D.~C.}\ \bibnamefont {Chrzan}}, \bibinfo {author} {\bibfnamefont {A.~M.}\ \bibnamefont {Minor}},\ and\ \bibinfo {author} {\bibfnamefont {A.}~\bibnamefont {Javey}},\ }\bibfield  {title} {\bibinfo {title} {Strain-engineered growth of two-dimensional materials},\ }\href {https://www.nature.com/articles/s41467-017-00516-5} {\bibfield  {journal} {\bibinfo  {journal} {Nature communications}\ }\textbf {\bibinfo {volume} {8}},\ \bibinfo {pages} {608} (\bibinfo {year} {2017})}\BibitemShut {NoStop}%
\bibitem [{\citenamefont {Srivastava}\ \emph {et~al.}(2017{\natexlab{a}})\citenamefont {Srivastava}, \citenamefont {Yadav}, \citenamefont {Rani},\ and\ \citenamefont {Ghosh}}]{srivastava2017controlled}%
  \BibitemOpen
  \bibfield  {author} {\bibinfo {author} {\bibfnamefont {P.~K.}\ \bibnamefont {Srivastava}}, \bibinfo {author} {\bibfnamefont {P.}~\bibnamefont {Yadav}}, \bibinfo {author} {\bibfnamefont {V.}~\bibnamefont {Rani}},\ and\ \bibinfo {author} {\bibfnamefont {S.}~\bibnamefont {Ghosh}},\ }\bibfield  {title} {\bibinfo {title} {Controlled doping in graphene monolayers by trapping organic molecules at the graphene--substrate interface},\ }\href {https://pubs.acs.org/doi/10.1021/acsami.6b13211} {\bibfield  {journal} {\bibinfo  {journal} {ACS Applied Materials \& Interfaces}\ }\textbf {\bibinfo {volume} {9}},\ \bibinfo {pages} {5375} (\bibinfo {year} {2017}{\natexlab{a}})}\BibitemShut {NoStop}%
\bibitem [{SM()}]{SM}%
  \BibitemOpen
  \bibfield  {title} {\bibinfo {title} {See \uppercase{S}upplemental \uppercase{M}aterial [link to be inserted by the publisher] for sample preperation, device fabrication, device characterization, charge transfer mechanism, method used for density functional theory calculations, schematic illustration of strain, impact of doping and strain on \uppercase{R}aman modes and determination of the \uppercase{R}aman 2d peak shift in our samples. \uppercase{SM} also contain \uppercase{R}efs. \cite{berry2013impermeability, sun2020limits, kresse1996efficiency, perdew1992atoms, perdew1996generalized, goumans2007structure, srivastava2017relativistic, bruna2014doping, ferrari2013raman, barrios2013pseudo}},\ }\href@noop {} {\ }\BibitemShut {NoStop}%
\bibitem [{\citenamefont {Berry}(2013)}]{berry2013impermeability}%
  \BibitemOpen
  \bibfield  {author} {\bibinfo {author} {\bibfnamefont {V.}~\bibnamefont {Berry}},\ }\bibfield  {title} {\bibinfo {title} {Impermeability of graphene and its applications},\ }\href {https://www.sciencedirect.com/science/article/abs/pii/S0008622313004880} {\bibfield  {journal} {\bibinfo  {journal} {Carbon}\ }\textbf {\bibinfo {volume} {62}},\ \bibinfo {pages} {1} (\bibinfo {year} {2013})}\BibitemShut {NoStop}%
\bibitem [{\citenamefont {Sun}\ \emph {et~al.}(2020)\citenamefont {Sun}, \citenamefont {Yang}, \citenamefont {Kuang}, \citenamefont {Stebunov}, \citenamefont {Xiong}, \citenamefont {Yu}, \citenamefont {Nair}, \citenamefont {Katsnelson}, \citenamefont {Yuan}, \citenamefont {Grigorieva} \emph {et~al.}}]{sun2020limits}%
  \BibitemOpen
  \bibfield  {author} {\bibinfo {author} {\bibfnamefont {P.}~\bibnamefont {Sun}}, \bibinfo {author} {\bibfnamefont {Q.}~\bibnamefont {Yang}}, \bibinfo {author} {\bibfnamefont {W.}~\bibnamefont {Kuang}}, \bibinfo {author} {\bibfnamefont {Y.}~\bibnamefont {Stebunov}}, \bibinfo {author} {\bibfnamefont {W.}~\bibnamefont {Xiong}}, \bibinfo {author} {\bibfnamefont {J.}~\bibnamefont {Yu}}, \bibinfo {author} {\bibfnamefont {R.~R.}\ \bibnamefont {Nair}}, \bibinfo {author} {\bibfnamefont {M.}~\bibnamefont {Katsnelson}}, \bibinfo {author} {\bibfnamefont {S.}~\bibnamefont {Yuan}}, \bibinfo {author} {\bibfnamefont {I.}~\bibnamefont {Grigorieva}}, \emph {et~al.},\ }\bibfield  {title} {\bibinfo {title} {Limits on gas impermeability of graphene},\ }\href {https://www.nature.com/articles/s41586-020-2070-x} {\bibfield  {journal} {\bibinfo  {journal} {Nature}\ }\textbf {\bibinfo {volume} {579}},\ \bibinfo {pages} {229} (\bibinfo {year} {2020})}\BibitemShut {NoStop}%
\bibitem [{\citenamefont {Kresse}\ and\ \citenamefont {Furthm{\"u}ller}(1996)}]{kresse1996efficiency}%
  \BibitemOpen
  \bibfield  {author} {\bibinfo {author} {\bibfnamefont {G.}~\bibnamefont {Kresse}}\ and\ \bibinfo {author} {\bibfnamefont {J.}~\bibnamefont {Furthm{\"u}ller}},\ }\bibfield  {title} {\bibinfo {title} {Efficiency of ab-initio total energy calculations for metals and semiconductors using a plane-wave basis set},\ }\href {https://www.sciencedirect.com/science/article/abs/pii/0927025696000080} {\bibfield  {journal} {\bibinfo  {journal} {Computational materials science}\ }\textbf {\bibinfo {volume} {6}},\ \bibinfo {pages} {15} (\bibinfo {year} {1996})}\BibitemShut {NoStop}%
\bibitem [{\citenamefont {Perdew}\ \emph {et~al.}(1992)\citenamefont {Perdew}, \citenamefont {Chevary}, \citenamefont {Vosko}, \citenamefont {Jackson}, \citenamefont {Pederson}, \citenamefont {Singh},\ and\ \citenamefont {Fiolhais}}]{perdew1992atoms}%
  \BibitemOpen
  \bibfield  {author} {\bibinfo {author} {\bibfnamefont {J.~P.}\ \bibnamefont {Perdew}}, \bibinfo {author} {\bibfnamefont {J.~A.}\ \bibnamefont {Chevary}}, \bibinfo {author} {\bibfnamefont {S.~H.}\ \bibnamefont {Vosko}}, \bibinfo {author} {\bibfnamefont {K.~A.}\ \bibnamefont {Jackson}}, \bibinfo {author} {\bibfnamefont {M.~R.}\ \bibnamefont {Pederson}}, \bibinfo {author} {\bibfnamefont {D.~J.}\ \bibnamefont {Singh}},\ and\ \bibinfo {author} {\bibfnamefont {C.}~\bibnamefont {Fiolhais}},\ }\bibfield  {title} {\bibinfo {title} {Atoms, molecules, solids, and surfaces: Applications of the generalized gradient approximation for exchange and correlation},\ }\href {https://journals.aps.org/prb/abstract/10.1103/PhysRevB.46.6671} {\bibfield  {journal} {\bibinfo  {journal} {Physical review B}\ }\textbf {\bibinfo {volume} {46}},\ \bibinfo {pages} {6671} (\bibinfo {year} {1992})}\BibitemShut {NoStop}%
\bibitem [{\citenamefont {Perdew}\ \emph {et~al.}(1996)\citenamefont {Perdew}, \citenamefont {Burke},\ and\ \citenamefont {Ernzerhof}}]{perdew1996generalized}%
  \BibitemOpen
  \bibfield  {author} {\bibinfo {author} {\bibfnamefont {J.~P.}\ \bibnamefont {Perdew}}, \bibinfo {author} {\bibfnamefont {K.}~\bibnamefont {Burke}},\ and\ \bibinfo {author} {\bibfnamefont {M.}~\bibnamefont {Ernzerhof}},\ }\bibfield  {title} {\bibinfo {title} {Generalized gradient approximation made simple},\ }\href {https://journals.aps.org/prl/abstract/10.1103/PhysRevLett.77.3865} {\bibfield  {journal} {\bibinfo  {journal} {Physical review letters}\ }\textbf {\bibinfo {volume} {77}},\ \bibinfo {pages} {3865} (\bibinfo {year} {1996})}\BibitemShut {NoStop}%
\bibitem [{\citenamefont {Goumans}\ \emph {et~al.}(2007)\citenamefont {Goumans}, \citenamefont {Wander}, \citenamefont {Brown},\ and\ \citenamefont {Catlow}}]{goumans2007structure}%
  \BibitemOpen
  \bibfield  {author} {\bibinfo {author} {\bibfnamefont {T.}~\bibnamefont {Goumans}}, \bibinfo {author} {\bibfnamefont {A.}~\bibnamefont {Wander}}, \bibinfo {author} {\bibfnamefont {W.~A.}\ \bibnamefont {Brown}},\ and\ \bibinfo {author} {\bibfnamefont {C.~R.~A.}\ \bibnamefont {Catlow}},\ }\bibfield  {title} {\bibinfo {title} {Structure and stability of the (001) $\alpha$-quartz surface},\ }\href {https://pubs.rsc.org/en/content/articlelanding/2007/cp/b701176h} {\bibfield  {journal} {\bibinfo  {journal} {Physical Chemistry Chemical Physics}\ }\textbf {\bibinfo {volume} {9}},\ \bibinfo {pages} {2146} (\bibinfo {year} {2007})}\BibitemShut {NoStop}%
\bibitem [{\citenamefont {Srivastava}\ \emph {et~al.}(2017{\natexlab{b}})\citenamefont {Srivastava}, \citenamefont {Arya}, \citenamefont {Kumar},\ and\ \citenamefont {Ghosh}}]{srivastava2017relativistic}%
  \BibitemOpen
  \bibfield  {author} {\bibinfo {author} {\bibfnamefont {P.~K.}\ \bibnamefont {Srivastava}}, \bibinfo {author} {\bibfnamefont {S.}~\bibnamefont {Arya}}, \bibinfo {author} {\bibfnamefont {S.}~\bibnamefont {Kumar}},\ and\ \bibinfo {author} {\bibfnamefont {S.}~\bibnamefont {Ghosh}},\ }\bibfield  {title} {\bibinfo {title} {Relativistic nature of carriers: Origin of electron-hole conduction asymmetry in monolayer graphene},\ }\href {https://journals.aps.org/prb/abstract/10.1103/PhysRevB.96.241407} {\bibfield  {journal} {\bibinfo  {journal} {Physical Review B}\ }\textbf {\bibinfo {volume} {96}},\ \bibinfo {pages} {241407} (\bibinfo {year} {2017}{\natexlab{b}})}\BibitemShut {NoStop}%
\bibitem [{\citenamefont {Bruna}\ \emph {et~al.}(2014)\citenamefont {Bruna}, \citenamefont {Ott}, \citenamefont {Ij\"{a}s}, \citenamefont {Yoon}, \citenamefont {Sassi},\ and\ \citenamefont {Ferrari}}]{bruna2014doping}%
  \BibitemOpen
  \bibfield  {author} {\bibinfo {author} {\bibfnamefont {M.}~\bibnamefont {Bruna}}, \bibinfo {author} {\bibfnamefont {A.~K.}\ \bibnamefont {Ott}}, \bibinfo {author} {\bibfnamefont {M.}~\bibnamefont {Ij\"{a}s}}, \bibinfo {author} {\bibfnamefont {D.}~\bibnamefont {Yoon}}, \bibinfo {author} {\bibfnamefont {U.}~\bibnamefont {Sassi}},\ and\ \bibinfo {author} {\bibfnamefont {A.~C.}\ \bibnamefont {Ferrari}},\ }\bibfield  {title} {\bibinfo {title} {Doping dependence of the raman spectrum of defected graphene},\ }\href {https://pubs.acs.org/doi/10.1021/nn502676g} {\bibfield  {journal} {\bibinfo  {journal} {Acs Nano}\ }\textbf {\bibinfo {volume} {8}},\ \bibinfo {pages} {7432} (\bibinfo {year} {2014})}\BibitemShut {NoStop}%
\bibitem [{\citenamefont {Ferrari}\ and\ \citenamefont {Basko}(2013)}]{ferrari2013raman}%
  \BibitemOpen
  \bibfield  {author} {\bibinfo {author} {\bibfnamefont {A.~C.}\ \bibnamefont {Ferrari}}\ and\ \bibinfo {author} {\bibfnamefont {D.~M.}\ \bibnamefont {Basko}},\ }\bibfield  {title} {\bibinfo {title} {Raman spectroscopy as a versatile tool for studying the properties of graphene},\ }\href {https://www.nature.com/articles/nnano.2013.46} {\bibfield  {journal} {\bibinfo  {journal} {Nature nanotechnology}\ }\textbf {\bibinfo {volume} {8}},\ \bibinfo {pages} {235} (\bibinfo {year} {2013})}\BibitemShut {NoStop}%
\bibitem [{\citenamefont {Barrios-Vargas}\ and\ \citenamefont {Naumis}(2013)}]{barrios2013pseudo}%
  \BibitemOpen
  \bibfield  {author} {\bibinfo {author} {\bibfnamefont {J.}~\bibnamefont {Barrios-Vargas}}\ and\ \bibinfo {author} {\bibfnamefont {G.~G.}\ \bibnamefont {Naumis}},\ }\bibfield  {title} {\bibinfo {title} {Pseudo-gap opening and dirac point confined states in doped graphene},\ }\href {https://www.sciencedirect.com/science/article/abs/pii/S0038109813001221} {\bibfield  {journal} {\bibinfo  {journal} {Solid state communications}\ }\textbf {\bibinfo {volume} {162}},\ \bibinfo {pages} {23} (\bibinfo {year} {2013})}\BibitemShut {NoStop}%
\bibitem [{\citenamefont {Joshi}\ \emph {et~al.}(2014)\citenamefont {Joshi}, \citenamefont {Carbone}, \citenamefont {Wang}, \citenamefont {Kravets}, \citenamefont {Su}, \citenamefont {Grigorieva}, \citenamefont {Wu}, \citenamefont {Geim},\ and\ \citenamefont {Nair}}]{joshi2014precise}%
  \BibitemOpen
  \bibfield  {author} {\bibinfo {author} {\bibfnamefont {R.}~\bibnamefont {Joshi}}, \bibinfo {author} {\bibfnamefont {P.}~\bibnamefont {Carbone}}, \bibinfo {author} {\bibfnamefont {F.-C.}\ \bibnamefont {Wang}}, \bibinfo {author} {\bibfnamefont {V.~G.}\ \bibnamefont {Kravets}}, \bibinfo {author} {\bibfnamefont {Y.}~\bibnamefont {Su}}, \bibinfo {author} {\bibfnamefont {I.~V.}\ \bibnamefont {Grigorieva}}, \bibinfo {author} {\bibfnamefont {H.}~\bibnamefont {Wu}}, \bibinfo {author} {\bibfnamefont {A.~K.}\ \bibnamefont {Geim}},\ and\ \bibinfo {author} {\bibfnamefont {R.~R.}\ \bibnamefont {Nair}},\ }\bibfield  {title} {\bibinfo {title} {Precise and ultrafast molecular sieving through graphene oxide membranes},\ }\href {https://www.science.org/doi/10.1126/science.1245711} {\bibfield  {journal} {\bibinfo  {journal} {science}\ }\textbf {\bibinfo {volume} {343}},\ \bibinfo {pages} {752} (\bibinfo {year} {2014})}\BibitemShut {NoStop}%
\bibitem [{\citenamefont {Tsetseris}\ and\ \citenamefont {Pantelides}(2014)}]{tsetseris2014graphene}%
  \BibitemOpen
  \bibfield  {author} {\bibinfo {author} {\bibfnamefont {L.}~\bibnamefont {Tsetseris}}\ and\ \bibinfo {author} {\bibfnamefont {S.}~\bibnamefont {Pantelides}},\ }\bibfield  {title} {\bibinfo {title} {Graphene: An impermeable or selectively permeable membrane for atomic species},\ }\href {https://www.sciencedirect.com/science/article/abs/pii/S0008622313009093} {\bibfield  {journal} {\bibinfo  {journal} {Carbon}\ }\textbf {\bibinfo {volume} {67}},\ \bibinfo {pages} {58} (\bibinfo {year} {2014})}\BibitemShut {NoStop}%
\bibitem [{\citenamefont {Bunch}\ \emph {et~al.}(2008)\citenamefont {Bunch}, \citenamefont {Verbridge}, \citenamefont {Alden}, \citenamefont {Van Der~Zande}, \citenamefont {Parpia}, \citenamefont {Craighead},\ and\ \citenamefont {McEuen}}]{bunch2008impermeable}%
  \BibitemOpen
  \bibfield  {author} {\bibinfo {author} {\bibfnamefont {J.~S.}\ \bibnamefont {Bunch}}, \bibinfo {author} {\bibfnamefont {S.~S.}\ \bibnamefont {Verbridge}}, \bibinfo {author} {\bibfnamefont {J.~S.}\ \bibnamefont {Alden}}, \bibinfo {author} {\bibfnamefont {A.~M.}\ \bibnamefont {Van Der~Zande}}, \bibinfo {author} {\bibfnamefont {J.~M.}\ \bibnamefont {Parpia}}, \bibinfo {author} {\bibfnamefont {H.~G.}\ \bibnamefont {Craighead}},\ and\ \bibinfo {author} {\bibfnamefont {P.~L.}\ \bibnamefont {McEuen}},\ }\bibfield  {title} {\bibinfo {title} {Impermeable atomic membranes from graphene sheets},\ }\href {https://pubs.acs.org/doi/10.1021/nl801457b} {\bibfield  {journal} {\bibinfo  {journal} {Nano letters}\ }\textbf {\bibinfo {volume} {8}},\ \bibinfo {pages} {2458} (\bibinfo {year} {2008})}\BibitemShut {NoStop}%
\bibitem [{\citenamefont {Khandelwal}\ \emph {et~al.}(2023)\citenamefont {Khandelwal}, \citenamefont {Srivastava}, \citenamefont {Nagaraja}, \citenamefont {Yadav}, \citenamefont {Tarafder},\ and\ \citenamefont {Ghosh}}]{khandelwal2023revealing}%
  \BibitemOpen
  \bibfield  {author} {\bibinfo {author} {\bibfnamefont {V.}~\bibnamefont {Khandelwal}}, \bibinfo {author} {\bibfnamefont {P.~K.}\ \bibnamefont {Srivastava}}, \bibinfo {author} {\bibfnamefont {S.}~\bibnamefont {Nagaraja}}, \bibinfo {author} {\bibfnamefont {P.}~\bibnamefont {Yadav}}, \bibinfo {author} {\bibfnamefont {K.}~\bibnamefont {Tarafder}},\ and\ \bibinfo {author} {\bibfnamefont {S.}~\bibnamefont {Ghosh}},\ }\bibfield  {title} {\bibinfo {title} {Revealing the microscopic picture of the charge transfer mechanism between graphene and dopant molecules},\ }\href {https://pubs.acs.org/doi/10.1021/acs.jpcc.3c03524} {\bibfield  {journal} {\bibinfo  {journal} {The Journal of Physical Chemistry C}\ }\textbf {\bibinfo {volume} {127}},\ \bibinfo {pages} {18466} (\bibinfo {year} {2023})}\BibitemShut {NoStop}%
\bibitem [{\citenamefont {Dong}\ \emph {et~al.}(2009)\citenamefont {Dong}, \citenamefont {Fu}, \citenamefont {Fang}, \citenamefont {Shi}, \citenamefont {Chen},\ and\ \citenamefont {Li}}]{dong2009doping}%
  \BibitemOpen
  \bibfield  {author} {\bibinfo {author} {\bibfnamefont {X.}~\bibnamefont {Dong}}, \bibinfo {author} {\bibfnamefont {D.}~\bibnamefont {Fu}}, \bibinfo {author} {\bibfnamefont {W.}~\bibnamefont {Fang}}, \bibinfo {author} {\bibfnamefont {Y.}~\bibnamefont {Shi}}, \bibinfo {author} {\bibfnamefont {P.}~\bibnamefont {Chen}},\ and\ \bibinfo {author} {\bibfnamefont {L.~J.}\ \bibnamefont {Li}},\ }\bibfield  {title} {\bibinfo {title} {Doping single-layer graphene with aromatic molecules},\ }\href {https://onlinelibrary.wiley.com/doi/10.1002/smll.200801711} {\bibfield  {journal} {\bibinfo  {journal} {Small}\ } (\bibinfo {year} {2009})}\BibitemShut {NoStop}%
\bibitem [{\citenamefont {Ren}(2004)}]{ren2004large}%
  \BibitemOpen
  \bibfield  {author} {\bibinfo {author} {\bibfnamefont {X.}~\bibnamefont {Ren}},\ }\bibfield  {title} {\bibinfo {title} {Large electric-field-induced strain in ferroelectric crystals by point-defect-mediated reversible domain switching},\ }\href {https://www.nature.com/articles/nmat1051} {\bibfield  {journal} {\bibinfo  {journal} {Nature materials}\ }\textbf {\bibinfo {volume} {3}},\ \bibinfo {pages} {91} (\bibinfo {year} {2004})}\BibitemShut {NoStop}%
\bibitem [{\citenamefont {Voggu}\ \emph {et~al.}(2008)\citenamefont {Voggu}, \citenamefont {Das}, \citenamefont {Rout},\ and\ \citenamefont {Rao}}]{voggu2008effects}%
  \BibitemOpen
  \bibfield  {author} {\bibinfo {author} {\bibfnamefont {R.}~\bibnamefont {Voggu}}, \bibinfo {author} {\bibfnamefont {B.}~\bibnamefont {Das}}, \bibinfo {author} {\bibfnamefont {C.~S.}\ \bibnamefont {Rout}},\ and\ \bibinfo {author} {\bibfnamefont {C.}~\bibnamefont {Rao}},\ }\bibfield  {title} {\bibinfo {title} {Effects of charge transfer interaction of graphene with electron donor and acceptor molecules examined using raman spectroscopy and cognate techniques},\ }\href {https://iopscience.iop.org/article/10.1088/0953-8984/20/47/472204} {\bibfield  {journal} {\bibinfo  {journal} {Journal of Physics: Condensed Matter}\ }\textbf {\bibinfo {volume} {20}},\ \bibinfo {pages} {472204} (\bibinfo {year} {2008})}\BibitemShut {NoStop}%
\bibitem [{\citenamefont {Das}\ \emph {et~al.}(2008)\citenamefont {Das}, \citenamefont {Pisana}, \citenamefont {Chakraborty}, \citenamefont {Piscanec}, \citenamefont {Saha}, \citenamefont {Waghmare}, \citenamefont {Novoselov}, \citenamefont {Krishnamurthy}, \citenamefont {Geim}, \citenamefont {Ferrari} \emph {et~al.}}]{das2008monitoring}%
  \BibitemOpen
  \bibfield  {author} {\bibinfo {author} {\bibfnamefont {A.}~\bibnamefont {Das}}, \bibinfo {author} {\bibfnamefont {S.}~\bibnamefont {Pisana}}, \bibinfo {author} {\bibfnamefont {B.}~\bibnamefont {Chakraborty}}, \bibinfo {author} {\bibfnamefont {S.}~\bibnamefont {Piscanec}}, \bibinfo {author} {\bibfnamefont {S.~K.}\ \bibnamefont {Saha}}, \bibinfo {author} {\bibfnamefont {U.~V.}\ \bibnamefont {Waghmare}}, \bibinfo {author} {\bibfnamefont {K.~S.}\ \bibnamefont {Novoselov}}, \bibinfo {author} {\bibfnamefont {H.~R.}\ \bibnamefont {Krishnamurthy}}, \bibinfo {author} {\bibfnamefont {A.~K.}\ \bibnamefont {Geim}}, \bibinfo {author} {\bibfnamefont {A.~C.}\ \bibnamefont {Ferrari}}, \emph {et~al.},\ }\bibfield  {title} {\bibinfo {title} {Monitoring dopants by raman scattering in an electrochemically top-gated graphene transistor},\ }\href {https://www.nature.com/articles/nnano.2008.67} {\bibfield  {journal} {\bibinfo  {journal} {Nature nanotechnology}\ }\textbf {\bibinfo {volume} {3}},\ \bibinfo {pages} {210} (\bibinfo
  {year} {2008})}\BibitemShut {NoStop}%
\bibitem [{\citenamefont {Childres}\ \emph {et~al.}(2013)\citenamefont {Childres}, \citenamefont {Jauregui}, \citenamefont {Park}, \citenamefont {Cao}, \citenamefont {Chen} \emph {et~al.}}]{childres2013raman}%
  \BibitemOpen
  \bibfield  {author} {\bibinfo {author} {\bibfnamefont {I.}~\bibnamefont {Childres}}, \bibinfo {author} {\bibfnamefont {L.~A.}\ \bibnamefont {Jauregui}}, \bibinfo {author} {\bibfnamefont {W.}~\bibnamefont {Park}}, \bibinfo {author} {\bibfnamefont {H.}~\bibnamefont {Cao}}, \bibinfo {author} {\bibfnamefont {Y.~P.}\ \bibnamefont {Chen}}, \emph {et~al.},\ }\bibfield  {title} {\bibinfo {title} {Raman spectroscopy of graphene and related materials},\ }\href {https://www.sciencedirect.com/science/article/abs/pii/B9780323908009002523?via%3Dihub} {\bibfield  {journal} {\bibinfo  {journal} {New developments in photon and materials research}\ }\textbf {\bibinfo {volume} {1}},\ \bibinfo {pages} {1} (\bibinfo {year} {2013})}\BibitemShut {NoStop}%
\bibitem [{\citenamefont {Kertesz}\ \emph {et~al.}(1982)\citenamefont {Kertesz}, \citenamefont {Vonderviszt},\ and\ \citenamefont {Hoffman}}]{kertesz1982change}%
  \BibitemOpen
  \bibfield  {author} {\bibinfo {author} {\bibfnamefont {M.}~\bibnamefont {Kertesz}}, \bibinfo {author} {\bibfnamefont {F.}~\bibnamefont {Vonderviszt}},\ and\ \bibinfo {author} {\bibfnamefont {R.}~\bibnamefont {Hoffman}},\ }\bibfield  {title} {\bibinfo {title} {Change of c--c bond length in layers of graphite upon charge transfer},\ }\href {https://link.springer.com/article/10.1557/PROC-20-141} {\bibfield  {journal} {\bibinfo  {journal} {MRS Online Proceedings Library (OPL)}\ }\textbf {\bibinfo {volume} {20}},\ \bibinfo {pages} {141} (\bibinfo {year} {1982})}\BibitemShut {NoStop}%
\bibitem [{\citenamefont {Pietronero}\ and\ \citenamefont {Str{\"a}ssler}(1981)}]{pietronero1981bond}%
  \BibitemOpen
  \bibfield  {author} {\bibinfo {author} {\bibfnamefont {L.}~\bibnamefont {Pietronero}}\ and\ \bibinfo {author} {\bibfnamefont {S.}~\bibnamefont {Str{\"a}ssler}},\ }\bibfield  {title} {\bibinfo {title} {Bond-length change as a tool to determine charge transfer and electron-phonon coupling in graphite intercalation compounds},\ }\href {https://journals.aps.org/prl/abstract/10.1103/PhysRevLett.47.593} {\bibfield  {journal} {\bibinfo  {journal} {Physical Review Letters}\ }\textbf {\bibinfo {volume} {47}},\ \bibinfo {pages} {593} (\bibinfo {year} {1981})}\BibitemShut {NoStop}%
\bibitem [{\citenamefont {Metzger}\ \emph {et~al.}(2010)\citenamefont {Metzger}, \citenamefont {R{\'e}mi}, \citenamefont {Liu}, \citenamefont {Kusminskiy}, \citenamefont {Castro~Neto}, \citenamefont {Swan},\ and\ \citenamefont {Goldberg}}]{metzger2010biaxial}%
  \BibitemOpen
  \bibfield  {author} {\bibinfo {author} {\bibfnamefont {C.}~\bibnamefont {Metzger}}, \bibinfo {author} {\bibfnamefont {S.}~\bibnamefont {R{\'e}mi}}, \bibinfo {author} {\bibfnamefont {M.}~\bibnamefont {Liu}}, \bibinfo {author} {\bibfnamefont {S.~V.}\ \bibnamefont {Kusminskiy}}, \bibinfo {author} {\bibfnamefont {A.~H.}\ \bibnamefont {Castro~Neto}}, \bibinfo {author} {\bibfnamefont {A.~K.}\ \bibnamefont {Swan}},\ and\ \bibinfo {author} {\bibfnamefont {B.~B.}\ \bibnamefont {Goldberg}},\ }\bibfield  {title} {\bibinfo {title} {Biaxial strain in graphene adhered to shallow depressions},\ }\href {https://www.scholars.northwestern.edu/en/publications/biaxial-strain-in-graphene-adhered-to-shallow-depressions} {\bibfield  {journal} {\bibinfo  {journal} {Nano letters}\ }\textbf {\bibinfo {volume} {10}},\ \bibinfo {pages} {6} (\bibinfo {year} {2010})}\BibitemShut {NoStop}%
\bibitem [{\citenamefont {Androulidakis}\ \emph {et~al.}(2015)\citenamefont {Androulidakis}, \citenamefont {Koukaras}, \citenamefont {Parthenios}, \citenamefont {Kalosakas}, \citenamefont {Papagelis},\ and\ \citenamefont {Galiotis}}]{androulidakis2015graphene}%
  \BibitemOpen
  \bibfield  {author} {\bibinfo {author} {\bibfnamefont {C.}~\bibnamefont {Androulidakis}}, \bibinfo {author} {\bibfnamefont {E.~N.}\ \bibnamefont {Koukaras}}, \bibinfo {author} {\bibfnamefont {J.}~\bibnamefont {Parthenios}}, \bibinfo {author} {\bibfnamefont {G.}~\bibnamefont {Kalosakas}}, \bibinfo {author} {\bibfnamefont {K.}~\bibnamefont {Papagelis}},\ and\ \bibinfo {author} {\bibfnamefont {C.}~\bibnamefont {Galiotis}},\ }\bibfield  {title} {\bibinfo {title} {Graphene flakes under controlled biaxial deformation},\ }\href {https://www.nature.com/articles/srep18219} {\bibfield  {journal} {\bibinfo  {journal} {Scientific reports}\ }\textbf {\bibinfo {volume} {5}},\ \bibinfo {pages} {18219} (\bibinfo {year} {2015})}\BibitemShut {NoStop}%
\bibitem [{\citenamefont {Zabel}\ \emph {et~al.}(2012)\citenamefont {Zabel}, \citenamefont {Nair}, \citenamefont {Ott}, \citenamefont {Georgiou}, \citenamefont {Geim}, \citenamefont {Novoselov},\ and\ \citenamefont {Casiraghi}}]{zabel2012raman}%
  \BibitemOpen
  \bibfield  {author} {\bibinfo {author} {\bibfnamefont {J.}~\bibnamefont {Zabel}}, \bibinfo {author} {\bibfnamefont {R.~R.}\ \bibnamefont {Nair}}, \bibinfo {author} {\bibfnamefont {A.}~\bibnamefont {Ott}}, \bibinfo {author} {\bibfnamefont {T.}~\bibnamefont {Georgiou}}, \bibinfo {author} {\bibfnamefont {A.~K.}\ \bibnamefont {Geim}}, \bibinfo {author} {\bibfnamefont {K.~S.}\ \bibnamefont {Novoselov}},\ and\ \bibinfo {author} {\bibfnamefont {C.}~\bibnamefont {Casiraghi}},\ }\bibfield  {title} {\bibinfo {title} {Raman spectroscopy of graphene and bilayer under biaxial strain: bubbles and balloons},\ }\href {https://pubs.acs.org/doi/10.1021/nl203359n} {\bibfield  {journal} {\bibinfo  {journal} {Nano letters}\ }\textbf {\bibinfo {volume} {12}},\ \bibinfo {pages} {617} (\bibinfo {year} {2012})}\BibitemShut {NoStop}%
\bibitem [{\citenamefont {Badger}(1935)}]{badger1935relation}%
  \BibitemOpen
  \bibfield  {author} {\bibinfo {author} {\bibfnamefont {R.~M.}\ \bibnamefont {Badger}},\ }\bibfield  {title} {\bibinfo {title} {The relation between the internuclear distances and force constants of molecules and its application to polyatomic molecules},\ }\href {https://pubs.aip.org/aip/jcp/article-abstract/3/11/710/203342/The-Relation-Between-the-Internuclear-Distances?redirectedFrom=fulltext} {\bibfield  {journal} {\bibinfo  {journal} {The Journal of Chemical Physics}\ }\textbf {\bibinfo {volume} {3}},\ \bibinfo {pages} {710} (\bibinfo {year} {1935})}\BibitemShut {NoStop}%
\bibitem [{\citenamefont {Kaupp}\ \emph {et~al.}(2017)\citenamefont {Kaupp}, \citenamefont {Danovich},\ and\ \citenamefont {Shaik}}]{kaupp2017chemistry}%
  \BibitemOpen
  \bibfield  {author} {\bibinfo {author} {\bibfnamefont {M.}~\bibnamefont {Kaupp}}, \bibinfo {author} {\bibfnamefont {D.}~\bibnamefont {Danovich}},\ and\ \bibinfo {author} {\bibfnamefont {S.}~\bibnamefont {Shaik}},\ }\bibfield  {title} {\bibinfo {title} {Chemistry is about energy and its changes: A critique of bond-length/bond-strength correlations},\ }\href {https://www.sciencedirect.com/science/article/abs/pii/S0010854517300292?via%3Dihub} {\bibfield  {journal} {\bibinfo  {journal} {Coordination Chemistry Reviews}\ }\textbf {\bibinfo {volume} {344}},\ \bibinfo {pages} {355} (\bibinfo {year} {2017})}\BibitemShut {NoStop}%
\bibitem [{\citenamefont {Newaz}\ \emph {et~al.}(2012)\citenamefont {Newaz}, \citenamefont {Puzyrev}, \citenamefont {Wang}, \citenamefont {Pantelides},\ and\ \citenamefont {Bolotin}}]{newaz2012probing}%
  \BibitemOpen
  \bibfield  {author} {\bibinfo {author} {\bibfnamefont {A.}~\bibnamefont {Newaz}}, \bibinfo {author} {\bibfnamefont {Y.~S.}\ \bibnamefont {Puzyrev}}, \bibinfo {author} {\bibfnamefont {B.}~\bibnamefont {Wang}}, \bibinfo {author} {\bibfnamefont {S.~T.}\ \bibnamefont {Pantelides}},\ and\ \bibinfo {author} {\bibfnamefont {K.~I.}\ \bibnamefont {Bolotin}},\ }\bibfield  {title} {\bibinfo {title} {Probing charge scattering mechanisms in suspended graphene by varying its dielectric environment},\ }\href {https://www.nature.com/articles/ncomms1740} {\bibfield  {journal} {\bibinfo  {journal} {Nature Communications}\ }\textbf {\bibinfo {volume} {3}},\ \bibinfo {pages} {734} (\bibinfo {year} {2012})}\BibitemShut {NoStop}%
\bibitem [{\citenamefont {Srivastava}\ and\ \citenamefont {Ghosh}(2015)}]{srivastava2015defect}%
  \BibitemOpen
  \bibfield  {author} {\bibinfo {author} {\bibfnamefont {P.~K.}\ \bibnamefont {Srivastava}}\ and\ \bibinfo {author} {\bibfnamefont {S.}~\bibnamefont {Ghosh}},\ }\bibfield  {title} {\bibinfo {title} {Defect engineering as a versatile route to estimate various scattering mechanisms in monolayer graphene on solid substrates},\ }\href {https://pubs.rsc.org/en/content/articlelanding/2015/nr/c5nr04293c} {\bibfield  {journal} {\bibinfo  {journal} {Nanoscale}\ }\textbf {\bibinfo {volume} {7}},\ \bibinfo {pages} {16079} (\bibinfo {year} {2015})}\BibitemShut {NoStop}%
\bibitem [{\citenamefont {Xia}\ \emph {et~al.}(2010)\citenamefont {Xia}, \citenamefont {Chen}, \citenamefont {Wiktor}, \citenamefont {Ferry},\ and\ \citenamefont {Tao}}]{xia2010effect}%
  \BibitemOpen
  \bibfield  {author} {\bibinfo {author} {\bibfnamefont {J.}~\bibnamefont {Xia}}, \bibinfo {author} {\bibfnamefont {F.}~\bibnamefont {Chen}}, \bibinfo {author} {\bibfnamefont {P.}~\bibnamefont {Wiktor}}, \bibinfo {author} {\bibfnamefont {D.}~\bibnamefont {Ferry}},\ and\ \bibinfo {author} {\bibfnamefont {N.}~\bibnamefont {Tao}},\ }\bibfield  {title} {\bibinfo {title} {Effect of top dielectric medium on gate capacitance of graphene field effect transistors: implications in mobility measurements and sensor applications},\ }\href {https://pubmed.ncbi.nlm.nih.gov/21090582/} {\bibfield  {journal} {\bibinfo  {journal} {Nano Letters}\ }\textbf {\bibinfo {volume} {10}},\ \bibinfo {pages} {5060} (\bibinfo {year} {2010})}\BibitemShut {NoStop}%
\bibitem [{\citenamefont {Ni}\ \emph {et~al.}(2010)\citenamefont {Ni}, \citenamefont {Ponomarenko}, \citenamefont {Nair}, \citenamefont {Yang}, \citenamefont {Anissimova}, \citenamefont {Grigorieva}, \citenamefont {Schedin}, \citenamefont {Blake}, \citenamefont {Shen}, \citenamefont {Hill} \emph {et~al.}}]{ni2010resonant}%
  \BibitemOpen
  \bibfield  {author} {\bibinfo {author} {\bibfnamefont {Z.}~\bibnamefont {Ni}}, \bibinfo {author} {\bibfnamefont {L.}~\bibnamefont {Ponomarenko}}, \bibinfo {author} {\bibfnamefont {R.}~\bibnamefont {Nair}}, \bibinfo {author} {\bibfnamefont {R.}~\bibnamefont {Yang}}, \bibinfo {author} {\bibfnamefont {S.}~\bibnamefont {Anissimova}}, \bibinfo {author} {\bibfnamefont {I.}~\bibnamefont {Grigorieva}}, \bibinfo {author} {\bibfnamefont {F.}~\bibnamefont {Schedin}}, \bibinfo {author} {\bibfnamefont {P.}~\bibnamefont {Blake}}, \bibinfo {author} {\bibfnamefont {Z.}~\bibnamefont {Shen}}, \bibinfo {author} {\bibfnamefont {E.}~\bibnamefont {Hill}}, \emph {et~al.},\ }\bibfield  {title} {\bibinfo {title} {On resonant scatterers as a factor limiting carrier mobility in graphene},\ }\href {https://pubs.acs.org/doi/abs/10.1021/nl101399r} {\bibfield  {journal} {\bibinfo  {journal} {Nano letters}\ }\textbf {\bibinfo {volume} {10}},\ \bibinfo {pages} {3868} (\bibinfo {year} {2010})}\BibitemShut {NoStop}%
\bibitem [{\citenamefont {Hong}\ \emph {et~al.}(2009)\citenamefont {Hong}, \citenamefont {Zou},\ and\ \citenamefont {Zhu}}]{hong2009quantum}%
  \BibitemOpen
  \bibfield  {author} {\bibinfo {author} {\bibfnamefont {X.}~\bibnamefont {Hong}}, \bibinfo {author} {\bibfnamefont {K.}~\bibnamefont {Zou}},\ and\ \bibinfo {author} {\bibfnamefont {J.}~\bibnamefont {Zhu}},\ }\bibfield  {title} {\bibinfo {title} {Quantum scattering time and its implications on scattering sources in graphene},\ }\href {https://journals.aps.org/prb/abstract/10.1103/PhysRevB.80.241415} {\bibfield  {journal} {\bibinfo  {journal} {Physical Review B—Condensed Matter and Materials Physics}\ }\textbf {\bibinfo {volume} {80}},\ \bibinfo {pages} {241415} (\bibinfo {year} {2009})}\BibitemShut {NoStop}%
\bibitem [{\citenamefont {Dean}\ \emph {et~al.}(2010)\citenamefont {Dean}, \citenamefont {Young}, \citenamefont {Meric}, \citenamefont {Lee}, \citenamefont {Wang}, \citenamefont {Sorgenfrei}, \citenamefont {Watanabe}, \citenamefont {Taniguchi}, \citenamefont {Kim}, \citenamefont {Shepard} \emph {et~al.}}]{dean2010boron}%
  \BibitemOpen
  \bibfield  {author} {\bibinfo {author} {\bibfnamefont {C.~R.}\ \bibnamefont {Dean}}, \bibinfo {author} {\bibfnamefont {A.~F.}\ \bibnamefont {Young}}, \bibinfo {author} {\bibfnamefont {I.}~\bibnamefont {Meric}}, \bibinfo {author} {\bibfnamefont {C.}~\bibnamefont {Lee}}, \bibinfo {author} {\bibfnamefont {L.}~\bibnamefont {Wang}}, \bibinfo {author} {\bibfnamefont {S.}~\bibnamefont {Sorgenfrei}}, \bibinfo {author} {\bibfnamefont {K.}~\bibnamefont {Watanabe}}, \bibinfo {author} {\bibfnamefont {T.}~\bibnamefont {Taniguchi}}, \bibinfo {author} {\bibfnamefont {P.}~\bibnamefont {Kim}}, \bibinfo {author} {\bibfnamefont {K.~L.}\ \bibnamefont {Shepard}}, \emph {et~al.},\ }\bibfield  {title} {\bibinfo {title} {Boron nitride substrates for high-quality graphene electronics},\ }\href {https://www.nature.com/articles/nnano.2010.172} {\bibfield  {journal} {\bibinfo  {journal} {Nature nanotechnology}\ }\textbf {\bibinfo {volume} {5}},\ \bibinfo {pages} {722} (\bibinfo {year} {2010})}\BibitemShut {NoStop}%
\bibitem [{\citenamefont {Gosling}\ \emph {et~al.}(2021)\citenamefont {Gosling}, \citenamefont {Makarovsky}, \citenamefont {Wang}, \citenamefont {Cottam}, \citenamefont {Greenaway}, \citenamefont {Patan{\`e}}, \citenamefont {Wildman}, \citenamefont {Tuck}, \citenamefont {Turyanska},\ and\ \citenamefont {Fromhold}}]{gosling2021universal}%
  \BibitemOpen
  \bibfield  {author} {\bibinfo {author} {\bibfnamefont {J.~H.}\ \bibnamefont {Gosling}}, \bibinfo {author} {\bibfnamefont {O.}~\bibnamefont {Makarovsky}}, \bibinfo {author} {\bibfnamefont {F.}~\bibnamefont {Wang}}, \bibinfo {author} {\bibfnamefont {N.~D.}\ \bibnamefont {Cottam}}, \bibinfo {author} {\bibfnamefont {M.~T.}\ \bibnamefont {Greenaway}}, \bibinfo {author} {\bibfnamefont {A.}~\bibnamefont {Patan{\`e}}}, \bibinfo {author} {\bibfnamefont {R.~D.}\ \bibnamefont {Wildman}}, \bibinfo {author} {\bibfnamefont {C.~J.}\ \bibnamefont {Tuck}}, \bibinfo {author} {\bibfnamefont {L.}~\bibnamefont {Turyanska}},\ and\ \bibinfo {author} {\bibfnamefont {T.~M.}\ \bibnamefont {Fromhold}},\ }\bibfield  {title} {\bibinfo {title} {Universal mobility characteristics of graphene originating from charge scattering by ionised impurities},\ }\href {https://www.nature.com/articles/s42005-021-00518-2} {\bibfield  {journal} {\bibinfo  {journal} {Communications Physics}\ }\textbf {\bibinfo {volume} {4}},\ \bibinfo {pages} {30}
  (\bibinfo {year} {2021})}\BibitemShut {NoStop}%
\bibitem [{\citenamefont {Metten}\ \emph {et~al.}(2016)\citenamefont {Metten}, \citenamefont {Froehlicher},\ and\ \citenamefont {Berciaud}}]{metten2016monitoring}%
  \BibitemOpen
  \bibfield  {author} {\bibinfo {author} {\bibfnamefont {D.}~\bibnamefont {Metten}}, \bibinfo {author} {\bibfnamefont {G.}~\bibnamefont {Froehlicher}},\ and\ \bibinfo {author} {\bibfnamefont {S.}~\bibnamefont {Berciaud}},\ }\bibfield  {title} {\bibinfo {title} {Monitoring electrostatically-induced deflection, strain and doping in suspended graphene using raman spectroscopy},\ }\href {https://iopscience.iop.org/article/10.1088/2053-1583/4/1/014004} {\bibfield  {journal} {\bibinfo  {journal} {2D Materials}\ }\textbf {\bibinfo {volume} {4}},\ \bibinfo {pages} {014004} (\bibinfo {year} {2016})}\BibitemShut {NoStop}%
\bibitem [{\citenamefont {Chen}\ \emph {et~al.}(2022)\citenamefont {Chen}, \citenamefont {Luo}, \citenamefont {Liang}, \citenamefont {Ling},\ and\ \citenamefont {Swan}}]{chen2022charge}%
  \BibitemOpen
  \bibfield  {author} {\bibinfo {author} {\bibfnamefont {Z.}~\bibnamefont {Chen}}, \bibinfo {author} {\bibfnamefont {W.}~\bibnamefont {Luo}}, \bibinfo {author} {\bibfnamefont {L.}~\bibnamefont {Liang}}, \bibinfo {author} {\bibfnamefont {X.}~\bibnamefont {Ling}},\ and\ \bibinfo {author} {\bibfnamefont {A.~K.}\ \bibnamefont {Swan}},\ }\bibfield  {title} {\bibinfo {title} {Charge separation in monolayer wse2 by strain engineering: implications for strain-induced diode action},\ }\href {https://pubs.acs.org/doi/10.1021/acsanm.2c03264} {\bibfield  {journal} {\bibinfo  {journal} {ACS Applied Nano Materials}\ }\textbf {\bibinfo {volume} {5}},\ \bibinfo {pages} {15095} (\bibinfo {year} {2022})}\BibitemShut {NoStop}%
\end{thebibliography}%
\end{document}